\documentclass{llncs}

\usepackage{makeidx} 
\usepackage{subfigure,epsfig,wrapfig}
\usepackage{amsmath,amscd,stmaryrd}
\usepackage{amssymb,amsmath,latexsym}
\usepackage{color}

\input xy
\xyoption{all} 

\newcommand{\figeps}[3]{ 
    \begin{figure}[h!tb]
        \def\epsfsize##1##2{#2##1}%
        \centerline{\epsfbox{#1.eps}}
        \caption{#3}
        \label{fig:#1}
     \end{figure}
}

\newtheorem{defi}{Def.}

\newtheorem{prop}{Prop.}

\newcommand{\srule}[3]{#1 \overset{#3}{\longrightarrow} #2}

\newcommand{\abb}[3]{#1 \colon #2 \rightarrow #3}

\begin{document}


\title{Pattern-based Model-to-Model Transformation: Long Version} 
\titlerunning{Pattern-Based M2M Transformation} 

\author{Juan de Lara\inst{1} \and Esther Guerra\inst{2}\footnote{Extended version of the paper from the Proc. of ICGT'08 (Leicester).}}
\authorrunning{Juan de Lara et al.} 

\tocauthor{ Juan de Lara (Universidad Aut\'onoma de Madrid), Esther
Guerra (Universidad Carlos III de Madrid)}

\institute{ Universidad Aut\'onoma de Madrid (Spain),
\email{jdelara@uam.es} \and Universidad Carlos III de Madrid
(Spain), \email{eguerra@inf.uc3m.es}}

\maketitle

\begin{abstract}

We present a new, high-level approach for the specification of model-to-model
transformations based on declarative patterns.
These are (atomic or composite) constraints on triple graphs declaring the
allowed or forbidden relationships between source and target models. In this way, a
transformation is defined by specifying a set of triple
graph constraints that should be satisfied by the result of the transformation.

The description of the transformation is then compiled into lower-level operational
mechanisms to perform forward or backward transformations, as
well as to establish mappings between two existent models. In this paper
we study one of such mechanisms based on the generation of operational triple graph
grammar rules. Moreover, we exploit deduction techniques at the
specification level to generate more specialized constraints (preserving the specification
semantics) reflecting pattern dependencies, from which additional rules can be derived.

This is an extended version of the paper submitted to ICGT'08, with additional definitions
and proofs.

\end{abstract}

\section{Introduction}\label{sec:introduction}

Model-Driven Development (MDD)~\cite{Volter06} is a software
engineering paradigm where models are the core asset. They are used 
to specify, simulate, test, verify and generate code
for the application to be built. Most of these activities include
the specification and execution of model transformations, some of
them between different languages. The transformation of a model
conformant to a meta-model into another one conformant to a
different meta-model is called model-to-model (M2M) transformation,
and is the topic of this paper.

There are two main approaches to M2M transformation: {\em
operational} and {\em declarative}. The first one is based on
rules or instructions that explicitly state how and when the elements of the target
model should be created starting from the elements of the source
one. In declarative approaches, a description of the mappings
between the source and target models is provided. This description
states the relation that should hold between two models rather than
how to create and link their elements. Declarative approaches are
higher-level than operational ones since they form a compact
description of a set of (operational) rules. In addition, they are
inherently bidirectional because they do not specify any causality. Thus,
they bring together in a single specification forward (i.e.
source-to-target) and backward (i.e. target-to-source)
transformations.

The state-of-the-art on declarative M2M transformation notations
includes a handful of languages (see Section~\ref{sec:related}).
However, sometimes they lack a formal foundation
and analysis techniques able to prove properties of the
transformation~\cite{QVT}. In other cases, specifications are
not fully declarative and may require a control mechanism
or defining a causality between existing elements and those 
to be created in a given relation~\cite{EhrigTGG,Schurr94}, 
introducing some degree of operationality.

The state-of-the-art on declarative M2M transformation notations
includes a handful of languages (see Section~\ref{sec:related}).
However, sometimes they lack a formal foundation
and analysis techniques able to prove properties of the
transformation~\cite{QVT}. In other cases, specifications are
not fully declarative and may require a control mechanism
or defining a causality between existing elements and those 
to be created in a given relation~\cite{EhrigTGG,Schurr94}, 
introducing some degree of operationality.

In this paper, we propose a purely declarative, formal approach to
M2M transformation based on {\em triple patterns} to express the relations between source and target models.
These are similar to graph constraints~\cite{Fundamentals} but for
triple graphs, made of two graphs related through an intermediate
one. Patterns can specify positive (the relation they declare must
hold) or negative information (the relation must not hold) and
can be constrained by positive and negative restrictions. This high-level
specification is compiled into lower-level mechanisms based on triple graph 
grammar operational rules~\cite{Schurr94} to achieve
forward and backward transformations, as well as to relate two existing
models. The compilation is performed in two steps. First, we employ
deduction rules to derive additional patterns that reflect pattern
dependencies and refine existing patterns with negative restrictions.
Then, a rule for
the chosen transformation direction is derived from each pattern. 

The advantages of our technique are the following. First, it is
purely declarative, based on patterns and constraints. This
contrasts with other declarative approaches (such as Triple Graph
Grammars (TGGs)~\cite{EhrigTGG,Schurr94,Konigs05}) where a causality has to be given
between the existing elements and the ones that have to be created.
As we consider and exploit interactions between patterns, these dependencies are automatically derived.
Second, it has a formal foundation that allows the study of the M2M
specification, in both declarative (i.e. patterns)
and operational (i.e. derived rules) formats. 
Finally, we have devised deduction
techniques, able to derive
semantic information from the very patterns. For example, having a
positive pattern demanding a certain structure and a negative one
forbidding its duplication allows generating two rules: one creating
the structure if it is not present, and another one reusing it if it
already exists.

\noindent {\bf Paper Organization}. 
Section~\ref{sec:tri_graphs} introduces triple graphs and patterns.
Section~\ref{sec:deduction} presents the deduction rules.
Section~\ref{sec:patterns} shows how to derive 
operational rules from a pattern specification. 
Section~\ref{sec:analysis} proposes some analysis techniques for
M2M specifications.
Section~\ref{sec:related}
relates the most prominent declarative approaches to M2M transformation
with our proposal 
and
Section~\ref{sec:conclusions} ends with the conclusions.


\section{Specifying Transformations: Triple Patterns}\label{sec:tri_graphs}


This section introduces the different kinds of triple patterns, their satisfiability and
the characteristics of the underlying operational mechanisms. These concepts rely on the notion of triple graph,
which we introduce first.

Triple graphs are made of two graphs related through an intermediate one. We can use any graph model
for these three graphs, from standard unattributed graphs $(V; E; \abb{s,t}{E}{V})$
to more complex attributed graphs (e.g. E-graphs~\cite{Fundamentals}).

\begin{defi}[Triple Graph] A triple graph $TrG=(G_s, G_c, G_t, \abb{cs}{V_{G_c}}{V_{G_s}},$ $\abb{ct}{V_{G_c}}{V_{G_t}})$
is made of two graphs $G_s$ and $G_t$ called source and target, related through the nodes of the correspondence graph $G_c$.
\end{defi}

Nodes in the correspondence graph $G_c$ have morphisms to nodes of the source and target
graphs. If $\exists m \in V_{G_c}$ s.t. $x \overset{cs} \longmapsfrom m \overset{ct} \longmapsto y$
we write $x \: rel \: y$.
Other kinds of mappings could be used as well, for example the simpler one in~\cite{EhrigTGG}, where the
correspondence functions are graph morphisms or the more complex one in~\cite{SOSYM} where the correspondence
functions can relate edges or be undefined.
We use the notation $TrG|_{x}$ (for $x \in \{s, t, c\}$) to refer to the $G_x$ component of $TrG$, and write
$\langle G_s, G_t \rangle$ for a triple graph with source and target graphs $G_s$ and $G_t$, and 
$\langle G_s, \emptyset, G_t \rangle$ for a triple graph with empty correspondence.

Next, we define triple graph morphisms as a triple of graph morphisms that preserve the correspondence functions.

\begin{defi}[Triple Graph Morphism] A triple graph morphism $f=(f_s, f_c, f_t):$ $TrG^1 \rightarrow TrG^2$
is made of three graph morphisms $\abb{f_x}{TrG^1|_x}{TrG^2|_x}$ (with $x=\{s, c, t\}$), where
$f_s|_V \circ cs^1 = cs^2 \circ f_c|_V$ and $f_t|_V \circ ct^1 = ct^2 \circ f_c|_V$.
\end{defi}

Source and target graphs can be typed by a type graph, or more in general by a {\em meta-model}, which includes
inheritance~\cite{deLara07}. In the latter case, we use the term {\em model} instead of graph.
Given meta-model $MM$, $L(MM)$ refers to the set of all valid models
conformant to (typed by) it. Similarly, we use the notion of meta-model triple~\cite{SOSYM} for the typing of triple graphs.

Triple patterns are similar to graph constraints~\cite{Fundamentals,HeckelW95}, 
but defined on triple graphs. We use them to describe the allowed and forbidden 
relationships between source and target
models. We consider both simple and composite patterns.

\begin{defi}[Pattern] \label{def:pattern}
Given triple injective morphism $\abb{q}{C}{Q}$ and sets 
$N_{Pre}=\{\abb{c_i}{Q}{C_i}\}_{i \in Pre}$,
$N_{Post}=\{\abb{c_j}{Q}{C_j}\}_{j \in Post}$ of 
negative pre- and post-conditions:

\begin{itemize}
\item $\underset{i \in Pre} \bigwedge \overleftarrow{N}(C_i) \Rightarrow P(Q) \underset{j \in Post} \bigwedge \overrightarrow{N}(C_j)$ 
is a simple pattern (S-Pattern).
\item $\underset{i \in Pre} \bigwedge \overleftarrow N(C_i) \wedge \overleftarrow P(C) \Rightarrow P(Q)
\underset{j \in Post} \bigwedge \overrightarrow{N}(C_j)$ is a composite pattern (C-Pattern).
\item $\overrightarrow N(C_j)$ is a negative pattern (N-Pattern).
\end{itemize}
\end{defi}

\noindent {\bf Remark.} The notation $\overleftarrow P(\cdot)$, $\overleftarrow N(\cdot)$ and $\overrightarrow N(\cdot)$ is just syntactic
sugar to indicate a positive pre-condition, a negative pre-condition or a negative post-condition.

Thus, an S-Pattern is made of a positive graph Q restricted by negative pre- and post-conditions ($Pre$ and $Post$ sets).
The intuition is that $Q$ should be present in triple graph $TrG$ whenever
no negative pre-condition $C_i$ is found; and if $Q$ is found, then no occurrence of
the negative post-conditions should be found. That is, while pre-conditions express restrictions for 
the pattern $Q$ to occur, post-conditions describe forbidden graphs.
A C-Pattern is an S-Pattern with an additional positive
pre-condition graph $C$. Thus an S-Pattern is a C-Pattern with $C$ and $q$ empty. Finally, an N-Pattern
is a C-Pattern where $C$ and $Q$ are empty and there is only one negative post-condition, 
forbidden to occur. 


A M2M specification is a conjunction of simple and composite patterns.

\begin{defi}[M2M Specification] A M2M specification $S=\bigwedge_{i \in I} P_i$ is a conjunction of
patterns, where each $P_i$ can be simple, composite or negative.
\end{defi}

\noindent {\bf Remark.} 
For technical reasons, we assume that initially in a specification 
only N-patterns have negative post-conditions. This is not a restriction, 
as any post-condition can be expressed as an N-pattern. In fact, a M2M specification 
is usually made of just N- and S-patterns, from which we automatically 
derive C-patterns with positive pre-conditions encoding pattern dependencies, and 
transform N-patterns into post-conditions for the other patterns (see Section~\ref{sec:deduction}).

\begin{wrapfigure}{r}{0.35\textwidth}
\centering
\vspace{-0.7cm}
\includegraphics[width=0.35\textwidth]{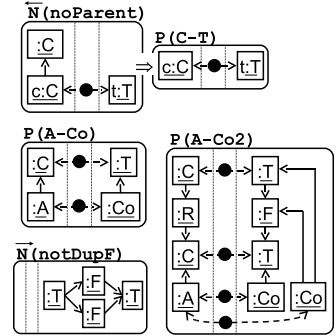}
\caption{M2M Specification.}
\label{fig:m2m}
\noindent 
\end{wrapfigure}

\noindent {\bf Example.} Fig.~\ref{fig:m2m} shows some patterns in an example M2M specification, inspired by 
the class to relational database transformation~\cite{QVT}. S-Pattern \texttt{C-T} states that a $C$ node 
(a class) that is not connected
to another one (i.e. it does not have a parent) should be related to a $T$ (table). 
S-Pattern \texttt{A-Co} states that a $C$ node connected to an $A$ (an attribute),
should be related to a $T$ with a $Co$ (column). Differently from TGGs, we don't need to
specify here a positive pre-condition stating that a relation between a $C$ and a $T$ should already exist. This
dependency is detected by the deduction rules we present in Section~\ref{sec:deduction}. S-Pattern \texttt{A-Co2}
specifies that in the case of two $C$ nodes connected through an $R$ (a directed relation),
the associated $T$ node of the source $C$ should have as foreign key ($F$ node) an attribute of the target
class. Finally, N-Pattern \texttt{notDupF} forbids two $F$s between two $T$s.

Next we define the satisfaction of a pattern. As S- and N-Patterns are special cases
of C-Patterns, it is enough to formulate C-Pattern satisfaction.

\begin{defi}[Pattern Satisfaction]\label{def:sat} Triple graph $TrG$ satisfies 
$CP= [\bigwedge_{i \in Pre} \overleftarrow N(C_i) \wedge \overleftarrow P(C) \Rightarrow P(Q)
\bigwedge_{j \in Post} \overrightarrow{N}(C_j)]$, 
written $TrG \models CP$, iff:

\begin{itemize} 
\item $CP$ is {\em forward satisfiable}, $TrG \models_F CP$: 
$[ \forall \abb{m^s}{P_s}{TrG}$ s.t. $(\forall i \in Pre$ s.t. $N^s_i \ncong P_s, \: \nexists \abb{n^s_i}{N^s_i}{TrG}$
with $m^s = n^s_i \circ a^s_i)$, $\exists \abb{m}{Q}{TrG}$ with $m \circ q^{s} = m^s$, s.t.
$\forall j \in Post$ $\nexists \abb{n_j}{C_j}{TrG}$ with $m = n_j \circ c_j]$, 

\item and $CP$ is {\em backwards satisfiable}, $TrG \models_B CP$: 
$[\forall \abb{m^t}{P_t}{TrG}$ s.t. $(\forall i \in Pre$ s.t. $N^t_i \ncong P_t, \: \nexists \abb{n^t_i}{N^t_i}{TrG}$
with $m^t = n^t_i \circ a^t_i)$, $\exists \abb{m}{Q}{TrG}$ with $m \circ q^{t} = m^t$, s.t.
$\forall j \in Post$ $\nexists \abb{n_j}{C_j}{TrG}$ with $m = n_j \circ c_j]$,
\end{itemize}

\noindent with $P_x = C +_{C|_x} Q|_x$, $N^x_i = C +_{C|_x} C_i|_x$  and
$N^x_i \overset{a^{x}_i} \longleftarrow P_x \overset {q^x} \longrightarrow Q$ $(x \in \{s, t\})$, 
see the left of Fig.~\ref{fig:sat_diag}\footnote{$A +_B C$ is the pushout object of $A$ and $C$ through $B$.
Similarly, $A \times_B C$ is the pullback object of $A$ and $C$ through C.}.

\end{defi}

\setlength{\unitlength}{0.7cm}
\begin{figure}[htb]
\begin{picture}(12, 4)

\put(0.5,4)
{\xymatrix@C=0.4cm@R=0.4cm{
C_i|_s\ar[d]_{d_i}\ar@{}[rd]|{P.O.}   &  & C|_s \ar@{^{(}->}[ld]\ar[rd]^{q_s}\ar[ll]_{b_s^i} \\
N^s_{i}\ar@/_2pc/[ddrr]|{/}_{n^s_{i}} & C\ar@{}[d]|{=}\ar[l]_{e^s_{i}}\ar[rd]^{c^s}\ar@{}[rr]|{P.O.} &  & Q|_s \ar[ld]_{p^s} & \\
       & & P_s\ar@{}[dl]_{=}\ar@/^1pc/[llu]|{a^s_i}\ar[d]_{m^s}\ar[r]^{q^s} & Q\ar@{}[dll]|<<<<<<<<{=}\ar@{}[d]|{=}\ar[r]^{c_{j}}\ar[ld]^{m} & C_{j}\ar@/^1pc/[lld]|{/}^{n_{j}} \\
       & & TrG & &
 }}

\put(10,-0.5) {\includegraphics[scale=0.35]{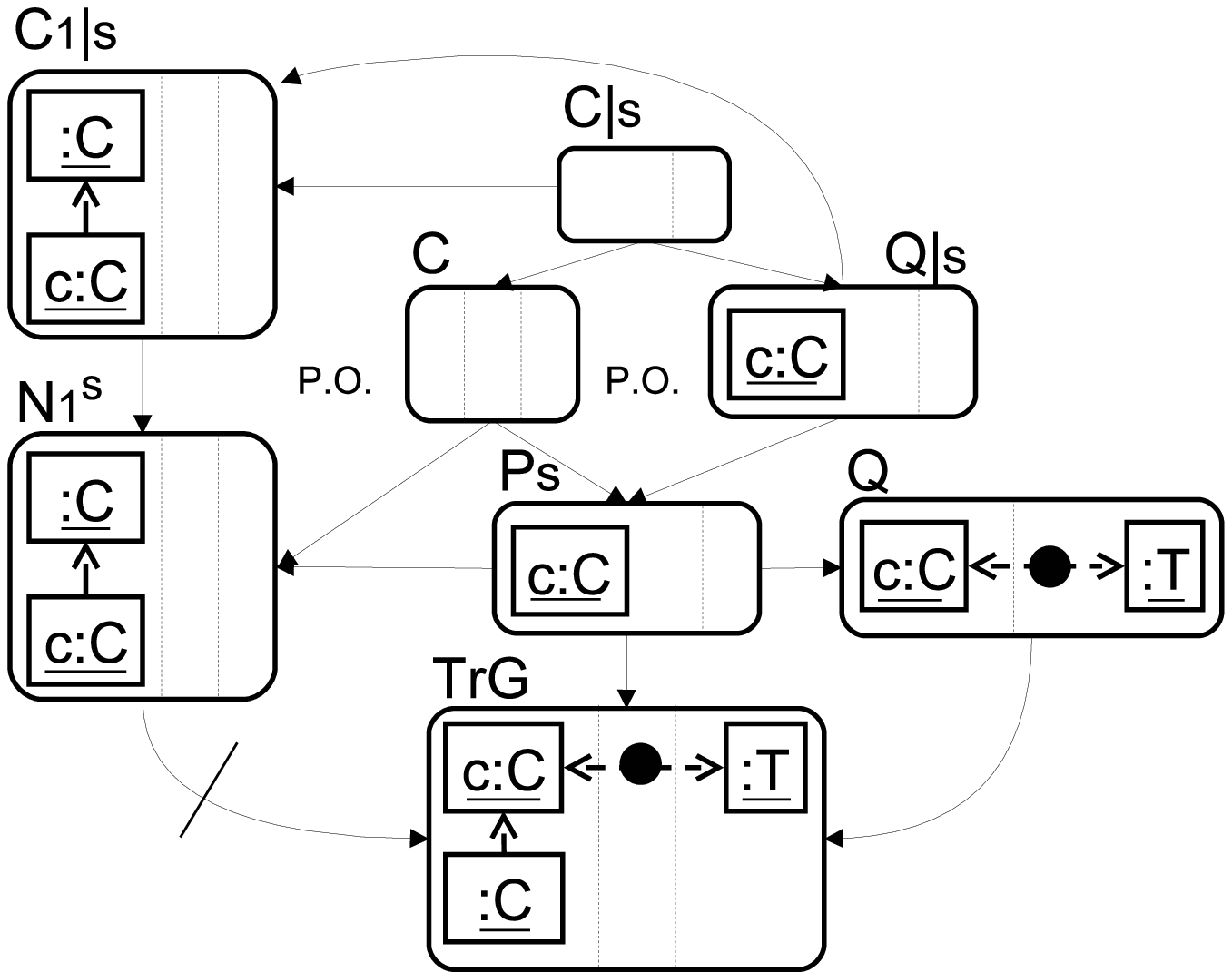}}
\end{picture}
\caption{Forward Satisfaction of Pattern (left). Forward Satisfaction Example (right).} \label{fig:sat_diag}
\end{figure}

\noindent {\bf Remark}. Morphisms $\abb{q^x}{P_x}{Q}$ ($x=\{s, t\}$)
uniquely exist due to the universal pushout property (as $ C|_x \hookrightarrow C \overset{q} \rightarrow Q = 
C|_x \overset{q_x} \rightarrow Q|_x \hookrightarrow Q$). For the same reason, 
$\abb{a^x_i}{P_x}{N^x_i}$ uniquely exist
(as $ C|_x \hookrightarrow C \overset{e^s_i} \rightarrow N^x_i = 
C|_x \overset{q_x} \rightarrow Q|_x \overset{c_i|_x} \rightarrow C_i|_x \overset{d_i} \rightarrow N^x_i$). Moreover, $b^i_x = c_i \circ q_x$.$\blacksquare$

C-Patterns have a universal quantification, therefore we split them into two
directed constraints. For this purpose we demand that, in forward satisfaction, for each occurrence of 
$P_s=Q|_s +_{C|_s} C = \langle Q|_s, C|_c, C|_t \rangle$ 
satisfying the
negative pre-conditions, an occurrence of $Q$ must be found satisfying the negative post-conditions, 
see the left of Fig.~\ref{fig:sat_diag}. 
A positive pattern graph $Q$ is satisfied either because no $m^s$ is found ({\em vacuous satisfaction}),
because $m^s$ and some negative pre-conditions are found ({\em negative satisfaction}), 
or because $m^s$ and $m$ are found and the negative pre- and post-conditions are not found ({\em positive satisfaction}).
Note that if the resulting directed negative pre-condition $N^x_i$ is isomorphic to $P_x$, then it is not taken
into account. This is needed as many pre-conditions express a restriction in either source or 
target but not on both.
In addition to forward satisfaction, similar conditions are demanded for the target graph (backwards satisfiability).
A graph satisfies specification $S$ if it satisfies all its patterns.


\noindent {\bf Example.} The right of Fig.~\ref{fig:sat_diag} shows the forward satisfaction of S-Pattern \texttt{C-T} by
a triple graph. $TrG \models_F C-T$ as there are two occurrences of $m^s$, the first one is shown in the figure 
(upper node $C$ in $TrG$) and is positively satisfied,
while the second (lower $C$) is negatively satisfied. We also have 
$TrG \models_B C-T$, as there is just one $m^t$, positively satisfied. This is because $N^t_1 \cong P_t$,
as we obtain a backward negative pre-condition with one $T$, which is isomorphic to $P_t$ and thus the negative condition
is not evaluated. Thus, $TrG \models C-T$. 

Please note that the specification does not explicitly state if a class
with a parent should be connected with a table or not. An additional pattern could describe such situation. The
forward operational mechanism, presented in Section~\ref{sec:patterns} does not add such table, as it minimally enforces the specification.

Starting from a specification $S$, lower level operational mechanisms are derived to perform forward ($\overrightarrow{S}$) and backward
transformations ($\overleftarrow{S}$), as well as to relate two existing models ($\overleftrightarrow{S}$). 
These mechanisms are described next.

\begin{defi}[Operational Mechanisms] Specification $S$ has the following associated operational transformations:
\begin{itemize}
\item {\bf Forward:} A function $\abb{\overrightarrow{S}}{V_S(MM_S)}{TrG}$ with domain 
$V_S(MM_S)=\{M_s \in L(MM_s) | \exists \langle M_s, X \rangle \models S\}$
s.t. $\forall M_{s} \in V_S(MM_S)$ 
$[\overrightarrow{S}(M_s) \models S] \wedge [\overrightarrow{S}(M_s)|_s \cong M_s]$.

\item {\bf Backwards:} A function $\abb{\overleftarrow{S}}{V_T(MM_T)}{TrG}$ with domain
$V_T(MM_T)=\{ M_t \in L(MM_t) | \exists \langle X, M_t \rangle \models S\}$
s.t. $\forall M_{t} \in V_T(MM_T)$
$[\overleftarrow{S}( M_t ) \models S] \wedge [\overleftarrow{S}( M_t )|_t \cong M_t]$.

\item {\bf Relating:} A function $\abb{\overleftrightarrow{S}}{V_{ST}(MM_s \times MM_t)}{TrG}$ with domain
$V_{ST}(MM_S \times MM_T)=\{ (M_s, M_t) | M_i \in L(MM_i) \wedge \exists \langle M_s, M_t \rangle \models S\}$ s.t.
$\forall (M_S, M_T) \in V_{ST}(MM_S, MM_T)$ 
$[\overleftrightarrow{S}(M_S, M_T) \models S] \wedge [\overleftrightarrow{S}(M_S, M_T)|_x \cong M_x]$ ($x=\{s, t\}$).

\end{itemize}
\end{defi}

The previous definitions are similar to the concept of {\em correct transformation} given in~\cite{Stevens},
but in addition we forbid modifying the source (resp. target) model in forward (resp. backwards) transformations. 

Next section presents some deduction rules, able to annotate patterns with dependencies, and also generate new ones.


\section{Deduction and Annotation Mechanisms for Patterns}\label{sec:deduction}


Next we present the deduction rules that we use to: (i) generate new patterns that take dependencies into account, which guide the 
order of pattern enforcement by the operational mechanism; (ii) enrich S-
and C-Patterns with pre- and post-conditions
derived from other patterns; and (iii) deduce positive information
from N-Patterns. For this purpose, we use two main operations: {\em deduction}, which infers new patterns, and
{\em annotation}, which makes dependencies among patterns explicit.

For example, from the specification in Fig.~\ref{fig:m2m}, the deduction rules generate a new pattern
to reflect the dependency between \texttt{C-T} and \texttt{A-Co} (to take into account whether a pair
$(C, T)$ is already related, before relating a pair $(A, Co)$). 
The deduction rules also add negative post-conditions derived from the \texttt{notDupF} N-pattern to 
the rest of patterns, and produce new patterns that reuse part of \texttt{notDupF} so that duplication of $F$ objects is not possible.

Most deduction rules are based on the
maximal intersection of two triple graphs, called {\em maximal intersection object} (MIO), which is defined next.

\begin{defi}[MIO] Given triple graphs $TrG_1$ and $TrG_2$, a maximal intersection (MI) is given by a span of injective 
morphisms $(TrG_1 \overset{m_1} \longleftarrow M  \overset{m_2} \longrightarrow TrG_2)$, 
s.t. $M \ncong \emptyset \wedge \nexists$ $M' \ncong M$ with $(TrG_1 \overset{m'_1} \longleftarrow M' \overset{m'_2} \longrightarrow TrG_2)$
and $\abb{m_{12}}{M}{M'}$ injective
s.t. the diagram to the left of Fig.~\ref{fig:MIO} commutes. Object $M$ is called MIO.
\end{defi}

MIOs are not unique, as the example to the right of Fig.~\ref{fig:MIO} shows: $M_1$ and $M_2$
are both MIOs, but not $M_3$ as $M_1$ is bigger.
The set of all MIs (resp. MIOs) of $TrG_1$ and $TrG_2$ is denoted by $MI(TrG_1, TrG_2)$ (resp. $MIO(TrG_1, TrG_2)$).

\setlength{\unitlength}{0.7cm}
\begin{figure}[htb]
\begin{picture}(12, 4.5)
\put(0,4.5)
{\xymatrix@C=0.3cm@R=0.5cm{
         &        & M'\ar[lldd]|{m'_{1}}\ar[rrdd]|{m'_{2}}\ar@{}[ddr]|{=}\ar@{}[ddl]|{=} & & \\
         &        & M\ar[u]|<<<{/}_{m_{12}}\ar[ld]|{m_1}\ar[rd]|{m_2}& & \\
m'_1(M')\ar@{}[dr]^{=}\ar@{^{(}->}@/_/[dr] & m_1(M)\ar@{_{(}->}[l]\ar@{_{(}->}[d] &  
& m_2(M)\ar@{^{(}->}[r]\ar@{_{(}->}[d] & m'_2(M')\ar@{_{(}->}@/^/[ld]\ar@{}[ld]_{=}\\
         & TrG_1  &  & TrG_2 & \\
 }}

\put(10.5,0) {\includegraphics[scale=0.35]{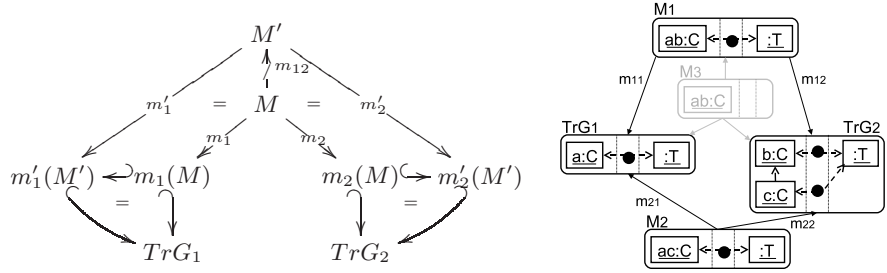}}
\end{picture}
\caption{Conditions for MIO (left). Example (right).} \label{fig:MIO}
\end{figure}

Patterns in a specification may have dependencies that induce a certain order of enforcement by
the operational mechanism. We make such dependencies explicit by annotating patterns with 
additional graphs, related to the positive graph $Q$. The dependency graphs are calculated by the intersection
of two patterns, and can be interpreted as restrictions that must not hold when the pattern is operationally enforced. 
The notion of annotated pattern is defined next.

\begin{defi}[Annotated Pattern] An annotated pattern $(P, \{\abb{n_k}{D_k}{Q}\}_{k \in K})$ contains
a pattern $P$ and a set of dependencies $D_k$ to $P$'s positive graph $Q$.
\end{defi}

Before presenting the deduction rules, we define an operation called {\em pre-condition weakening} (PW), which 
tests whether the positive graph of a C-Pattern is
included in another one, and then adds the negative pre-conditions from the former to the latter. 

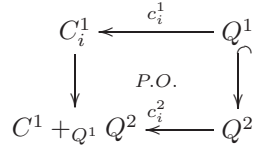
\begin{wrapfigure}{r}{0.35\textwidth}
\centering
{\xymatrix@C=0.9cm@R=0.7cm{
C^1_i\ar[d] & Q^1 \ar[l]_{c^1_i} \ar@{^{(}->}[d]\ar@{}[ld]|{P.O.}\\
C^1+_{Q^1}Q^2 & Q^2 \ar[l]_{c^2_i}
 }}
\caption{PW: Transferring the Negative Pre-Conditions.} \label{fig:PW}
\noindent 
\end{wrapfigure}

\begin{defi}[PW]\label{def:PW} Given
$\bigwedge_{k \in \{1, 2\}}[ \bigwedge_{i \in Pre^k} \overleftarrow{N}(C^k_i) \Rightarrow 
P(Q^k)]$ with $Q^1 \hookrightarrow Q^2$, $Pre^{11} \subseteq Pre^1$ and function $\abb{sub}{Pre^2}{Pre^{11}}$
surjective s.t. $\forall C^2_i \in Pre^2$, $\exists sub(C^2_i) \rightarrow C^2_i$ injective; the PW operation results in
$[ \bigwedge_{i \in Pre^1} \overleftarrow{N}(C^1_i) \Rightarrow 
P(Q^1)] \wedge 
[ \bigwedge_{i \in Pre^2} \overleftarrow{N}(C^2_i)$ $\bigwedge_{i \in Pre^1-Pre^{11}}$ $\overleftarrow{N}(C^1_i +_{Q^1} Q^2) \Rightarrow 
P(Q^2)]$, see Fig.~\ref{fig:PW}
\end{defi}

\noindent {\bf Remark.} The specification resulting from PW is not equivalent to the original one. The
second pattern is added negative pre-conditions, so that it is satisfiable by more graphs, namely by
those in which $\exists \abb{n^i}{C_i^1+_{Q^1}Q^2}{TrG}$ (injective), as then $Q^2$ is not forced to occur. However, we
use this operation to make coherent a specification: as an occurrence of the second pattern implies an occurrence of
the first, by adding the negative pre-conditions we ensure that a positive satisfaction of the second implies
a positive satisfaction of the first.

\noindent {\bf Example.} As S-Pattern \texttt{C-T} is included in \texttt{A-Co}, PW adds the negative
restriction $\overleftarrow N(noParent)$ from the former to the latter. The resulting pattern is shown to the right of Fig.~\ref{fig:S_Ded}
(second row, to the left).

Next, we show some deduction rules that preserve the specification semantics. 
We first present the deduction rule for two S-Patterns called {\em S-Deduction} and its annotation
mechanism $SA(\_\:,\_)$. S-Deduction creates a new pattern handling an intersection of
two S-Patterns, while the annotation mechanism adds such intersections dependencies
to the two original patterns. Its correctness proof is shown in the appendix.

\begin{prop}[S-Deduction]\label{prop:sp_ded} 
From $\bigwedge_{k \in \{1, 2\}} [ \bigwedge_{i \in Pre^k} \overleftarrow{N} (C^k_i) \Rightarrow 
P(Q^k)]$, we deduce the new patterns
$\bigwedge_{M \in MIO(Q^1, Q^2)} [ \bigwedge_{i \in Pre^1 \cup Pre^2} \overleftarrow{N}(C'_i) \wedge
 \overleftarrow{P}(M) \Rightarrow P(Q^1 +_{M} Q^2)]$,
where the $C'_i$ are calculated as shown to the left of Fig.~\ref{fig:S_Ded}.
\end{prop}

\setlength{\unitlength}{0.7cm}
\begin{figure}[htb]
\begin{picture}(12, 6.5)
\put(0,4.5)
{\xymatrix@C=0.2cm@R=0.5cm{
                         & M \ar[ld]|{m_1}\ar[rd]|{m_2}\ar@{}[dd]|{P.O.}& & \\
Q^1\ar[dd]_{c^1_i}\ar[rd] &                & Q^2\ar[dd]^{c^2_j}\ar[ld]  \\
                         & Q^1+_M Q^2\ar@<-2ex>[d]\ar@<2ex>[d]\ar@{}[rd]|{P.O.}\ar@{}[ld]|{P.O.} & \\
C^1_i\ar[r] & C'_i \:\:\: C'_j & C^2_j\ar[l]
 }}

\put(6,0) {\includegraphics[scale=0.35]{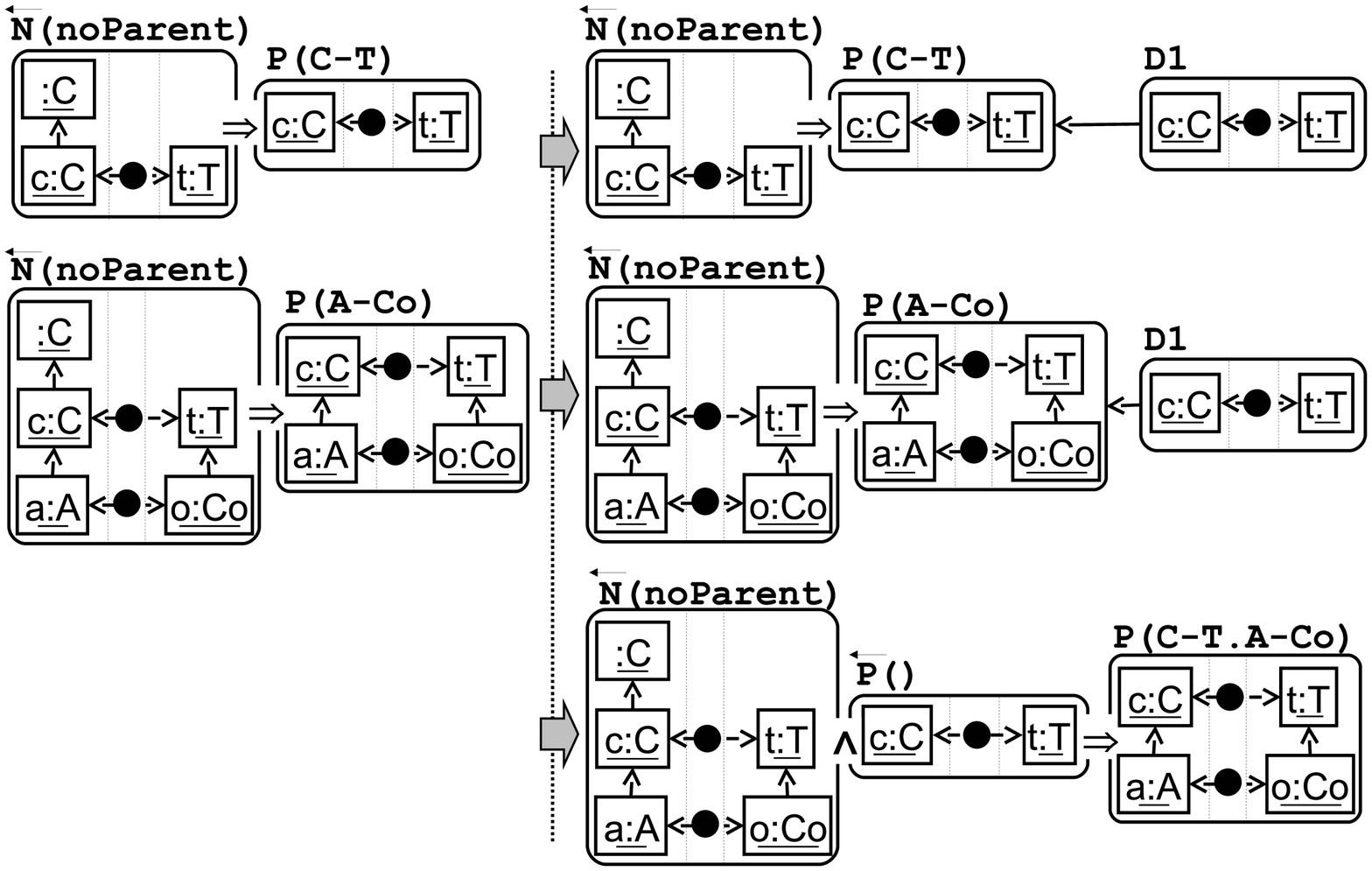}}
\end{picture}
\caption{Negative Pre-Conditions in S-Deduction (left). S-Annotation Example (right).} \label{fig:S_Ded}
\end{figure}


\begin{defi}[S-Annotation]\label{def:SA} Given two annotated S-Patterns $(P_i, D_i)$ with positive graphs $Q^i$:
$SA((P_1, D_1), (P_2, D_2))=\{ (P_i, $ $D_i \bigcup_{M \in MIO(Q^1, Q^2)} \{M \rightarrow Q^i\}) \}_{i=1, 2}$
$\bigcup_{M \in MIO(Q^1, Q^2)} \{(SD(P_1, P_2, M), \emptyset )\}$, with $SD(P_1, P_2, M)$ the resulting pattern 
from applying S-Deduction using $M$.
\end{defi}


\noindent {\bf Example.} The right of Fig.~\ref{fig:S_Ded} shows an example of S-Annotation, where the newly generated
pattern (bottom right) considers the fact that the relation demanded by pattern \texttt{C-T} may already exist. 
The procedure generates two ismorphic negative pre-conditions, so that one can be eliminated.
The added 
dependencies ($D_1$) ensure that the first and second patterns will only be enforced by the operational mechanisms when 
no occurrence of $D_1$ is found. As we will see later, this makes the TGG operational rules generated for the first two patterns
mutually exclusive with the one of the third, as well as confluent. Moreover, the rule for the third pattern will be able to reuse the
structure created by the rule of the first.

C-Deduction and its annotation mechanism $CA(\_\:,\_)$ 
are generalizations of the S- case, with the difference that the new pattern integrates
in its pre-condition the gluing of the original patterns pre-conditions.

\begin{figure}[htb]
\centering
\makebox{\xymatrix@C=0.8cm@R=0.8cm{
                &     & M^c \ar@{^{(}->}[d]\ar@/_1pc/[lld]|{mc^1}\ar@/^1pc/[rrd]|{mc^2} \\
C^1\ar[r]^{c^1}\ar[rd]|{pc^1} & Q^1\ar@/_3pc/[rddd] & M\ar@{}[dd]|{P.O.}\ar[l]_{m^1}\ar[r]^{m^2}\ar[ld]|{pm^1}\ar[rd]|{pm^2} & Q^2\ar@/^3pc/[lddd] & C^2\ar[l]_{c^2}\ar[ld]|{pc^2} \\
                & P^1\ar[u]|{pq^1}\ar[rd]|{p^1} &                        & P^2\ar[u]|{pq^2}\ar[ld]|{p^2} \\
                &       & P^c\ar[d] \\
                &       & Q^1+_M Q^2
 }}
\caption{C-Deduction.} \label{fig:C_Ded}
\end{figure}
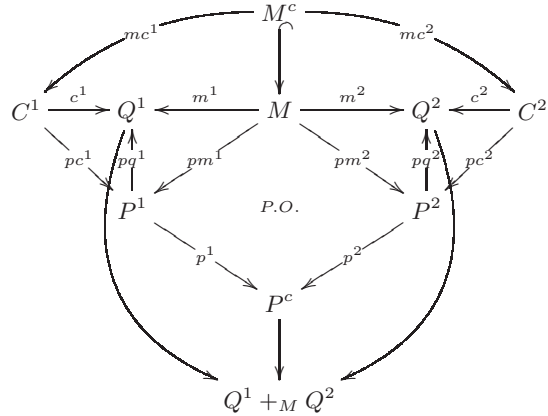

\begin{prop}[C-Deduction]\label{prop:cp_ded} 
From $\bigwedge_{k \in \{1, 2\}} [ \bigwedge_{i \in Pre^k} \overleftarrow{N} (C^k_i) \wedge \overleftarrow{P}(C^k) \Rightarrow 
P(Q^k)]$, we deduce the new pattern
$\bigwedge_{M \in MIO(Q_1, Q_2)} [ \bigwedge_{i \in Pre^1 \cup Pre^2} \overleftarrow{N}(C'_i) \wedge
 \overleftarrow{P}(P^c) \Rightarrow P(Q^1 +_{M} Q^2)]$,
where the $C'_i$ are calculated as shown to the left of Fig.~\ref{fig:S_Ded}, and $P^c$ is calculated as shown in Fig.~\ref{fig:C_Ded}.
In this diagram, $M^c$ is the subgraph of $M$ s.t. $(C^1 \overset{mc^1}\longleftarrow M^c \overset{mc^2}\longrightarrow C^2) \in
MI(C^1, C^2)$,
squares $pc^k \circ mc^k = pm^k \circ (M^c \hookrightarrow M) $ are P.O. and morphisms
$\abb{pq^k}{P^k}{Q^k}$ uniquely exist due to the P.O. universal property. For the same reason
$P^c \rightarrow Q^1+_M Q^2$ uniquely exists.
\end{prop}

\setlength{\unitlength}{0.7cm}

{\em Proof.} In appendix.$\blacksquare$

The annotation procedure for C-Deduction is analogous to the one for S-Deduction.


Next deduction rule is used to take into consideration the interaction of N-patterns, which express unconditional
negative constraints, with other patterns.

\begin{prop}[N-Deduction] \label{prop:N}
$[ \bigwedge_{i \in Pre} \overleftarrow{N}(C_i) \Rightarrow 
P(Q) \bigwedge_{j \in Post} \overrightarrow{N}(C_j)] \wedge [\overrightarrow N(C_N)]$ is equivalent to
$[ \bigwedge_{i \in Pre} \overleftarrow{N}(C_i) \Rightarrow 
P(Q) \bigwedge_{j \in Post} \overrightarrow{N}(C_j) \bigwedge_{C_{r} \in RS} \overrightarrow N(C_r)]$
with $RS=\{\abb{r^n}{Q}{C_{r}}\}_{C_r \in PO(MI(Q, C_N))}$ 
and $PO(MI(Q, C_N))$ is the set of pushout objects of all spans in $MI(Q, C_N)$.
\end{prop}

\noindent {\em Proof(Sketch).} We have related $C_N$ in all possible (maximal)
ways with $Q$, which is given by the pushout of each span in $MI(Q, C_N)$. This is similar to
the procedure to convert a graph constraint into a post-condition~\cite{Fundamentals,HeckelW95}. $\blacksquare$

\noindent {\bf Remark.} Removing $\overrightarrow N(C_N)$ does not yield an equivalent specification, as
e.g. a graph with no occurrence of $Q$ is allowed to have an occurrence of $C_N$. Note however that
we will delete N-Patterns when generating the TGG operational rules, as these by construction cannot generate
any forbidden pattern.

\figeps{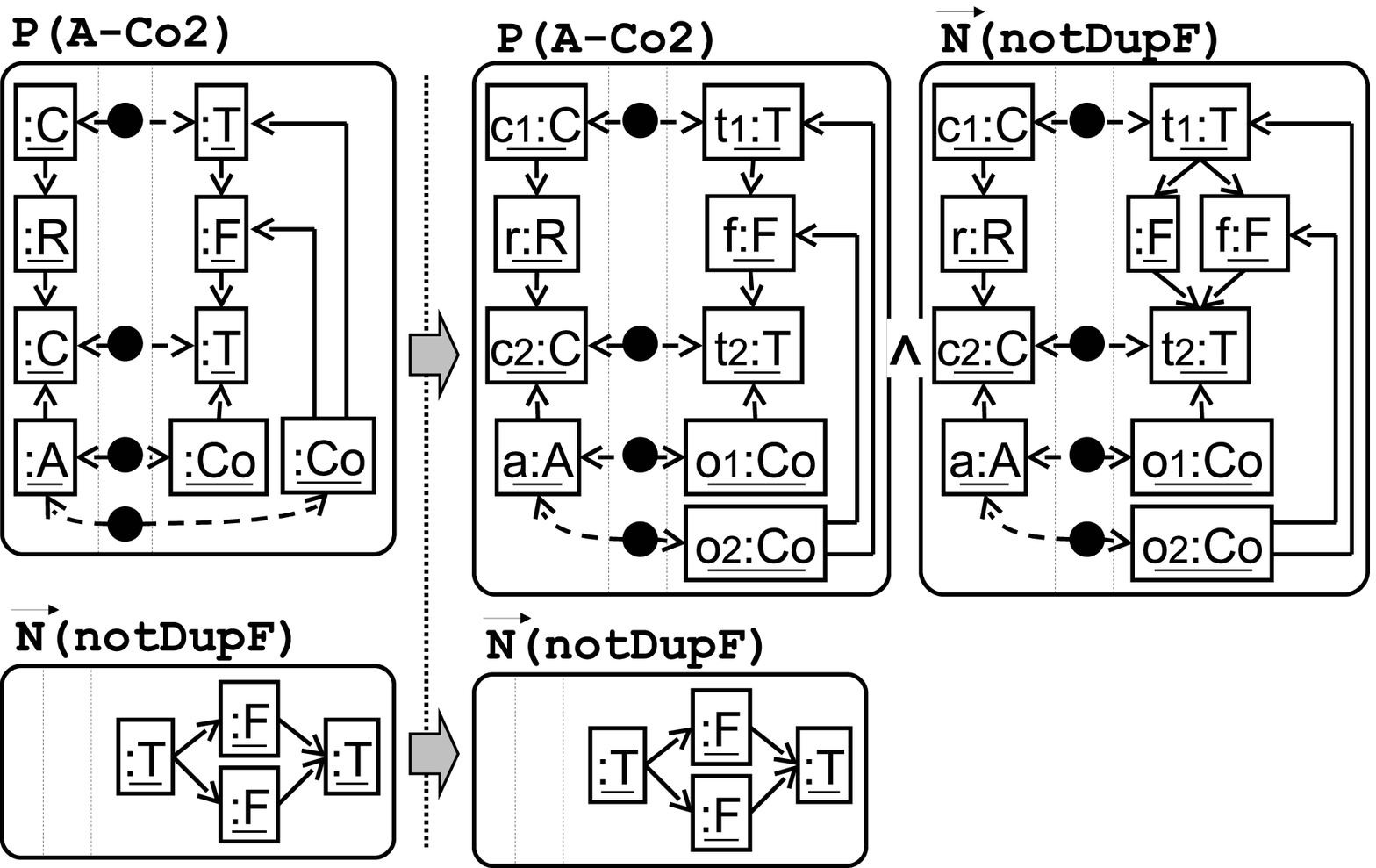}{0.4}{N-Deduction Example.}


\noindent {\bf Example.} Fig.~\ref{fig:SNDed_example} shows how N-Pattern \texttt{notDupF} induces
a negative constraint on S-Pattern \texttt{A-Co2}, resulting in the S-Pattern to its right. There are
two isomorphic MIOs (both made of two $T$s and one $F$) resulting in two isomorphic negative
constraints, so that one is eliminated.

The following deduction rule detects N-Patterns that forbid a repetition
of structures and generates a positive pattern that reuses such structure. First, we define
the completion of a triple graph $M$ with respect
to a graph $T$ such that $M \hookrightarrow T$. The completion adds to $M|_t$ all elements that are related to 
elements of $M|_s$ and belong to $T-M$, and similar for source elements. In addition, completion includes all
unrelated elements of $T$.

\begin{defi}[Completion] $C(M, T)= G$ iff $G$ is the smallest graph s.t. 
$M \hookrightarrow G \hookrightarrow T \wedge (\forall n \in V_{G|_s},  
\nexists m \in V_{T|_t}-V_{G|_t} s.t. \: n \: rel \: m) \wedge
(\forall x \in V_{G|_t},  
\nexists y \in V_{T|_s}-V_{G|_s} s.t. \: y \: rel \: x) \wedge (\nexists z \in (V_{T|_s} \cup V_{T|_t})-(V_{G|_s} 
\cup V_{G|_t})$ s.t. z is unrelated$)$. $G$ also contains all edges of $T$ with source and target in nodes of $G$.
\end{defi}

\figeps{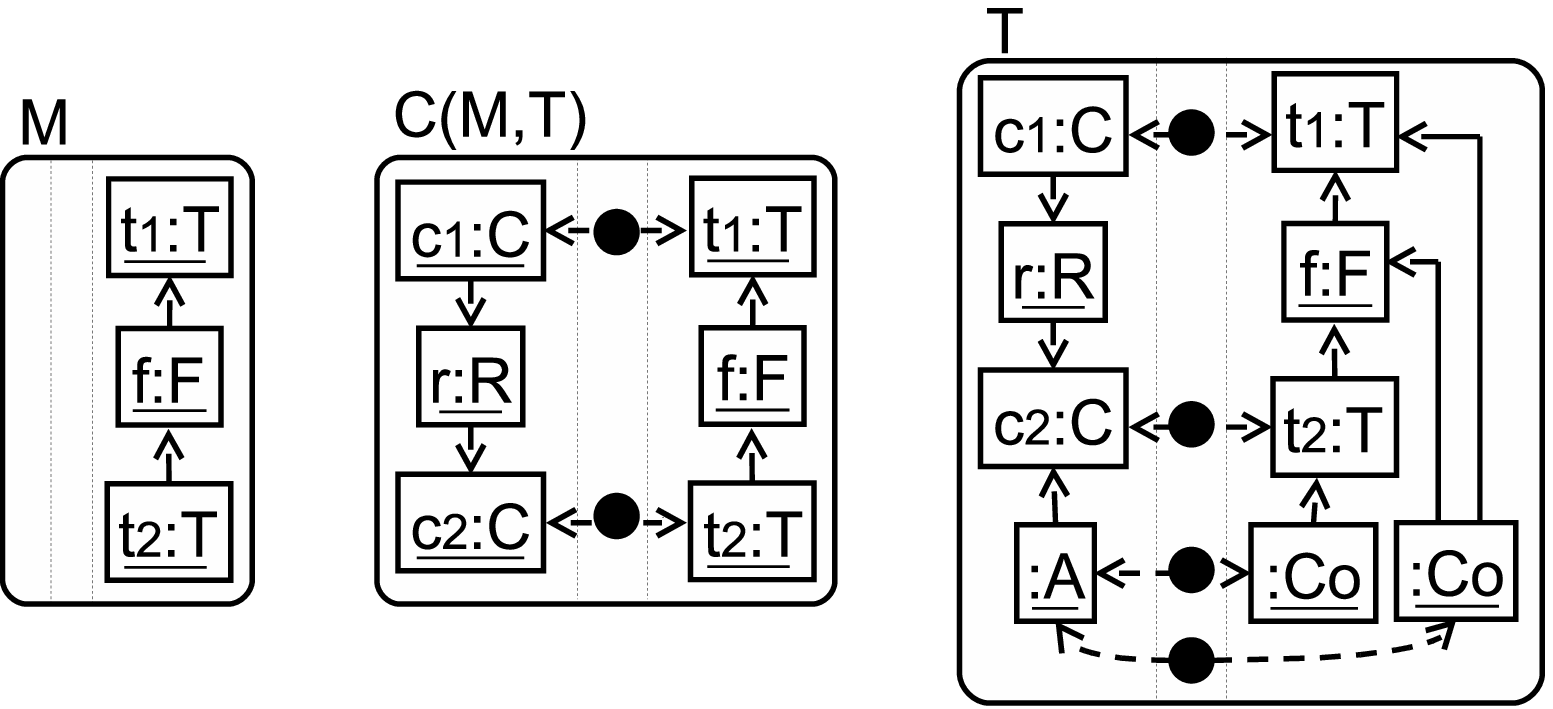}{0.4}{Example of Completion.}

\noindent {\bf Example.} Fig.~\ref{fig:completion} shows an example of completion, where
graph \texttt{M} is completed with respect to graph \texttt{T}, yielding graph
\texttt{C(M, T)}. Note that $M \hookrightarrow C(M, T) \hookrightarrow T$.

\begin{prop}[NP-Deduction] $[\bigwedge_{i \in Pre}
\overleftarrow N(C_i) \Rightarrow P(Q)] \wedge [\overrightarrow N(S)]$, with $S$ the pushout
of two isomorphic graphs $S_{1} \cong S_{2}$ and $S_1 \in MIO(Q, S)$, is equivalent 
to $[\bigwedge_{i \in Pre}$
$\overleftarrow N(C_i) \Rightarrow P(Q)] \wedge [\overrightarrow N(S)] \wedge 
[\bigwedge_{i \in Pre}
\overleftarrow N(C_i) \wedge \overleftarrow P(C(S_1, Q)) \Rightarrow P(Q)]$.
\end{prop}

\noindent {\em Proof.} $C(S_1, Q) \hookrightarrow Q$, thus $[\overleftarrow P(C(S_1, Q)) \Rightarrow P(Q)]$
is subsumed by $P(Q)$. $\blacksquare$

The NP-Deduction rule has an associated annotation rule $NP(\_\:,\_)$, which adds a dependency to the S-Pattern
equal to the positive pre-condition of the newly generated pattern.

\begin{defi}[NP-Annotation]\label{def:NA} $NP((\bigwedge_{i \in Pre} \overleftarrow N(C_i) \Rightarrow P(Q), D), 
(\overrightarrow N(S), D'))=\{(\bigwedge_{i \in Pre} \overleftarrow N(C_i) \Rightarrow P(Q), D \cup \{C(S_1, Q)\}), (\overrightarrow N(S), D'),
(\bigwedge_{i \in Pre} \overleftarrow N(C_i) \wedge \overleftarrow P(C(S_1, Q)) \Rightarrow P(Q), \emptyset )\}$, with $S$ the P.O.
of two isomorphic graphs $S_{1} \cong S_{2}$ and $S_1 \in MIO (S, Q)$.
\end{defi}

\figeps{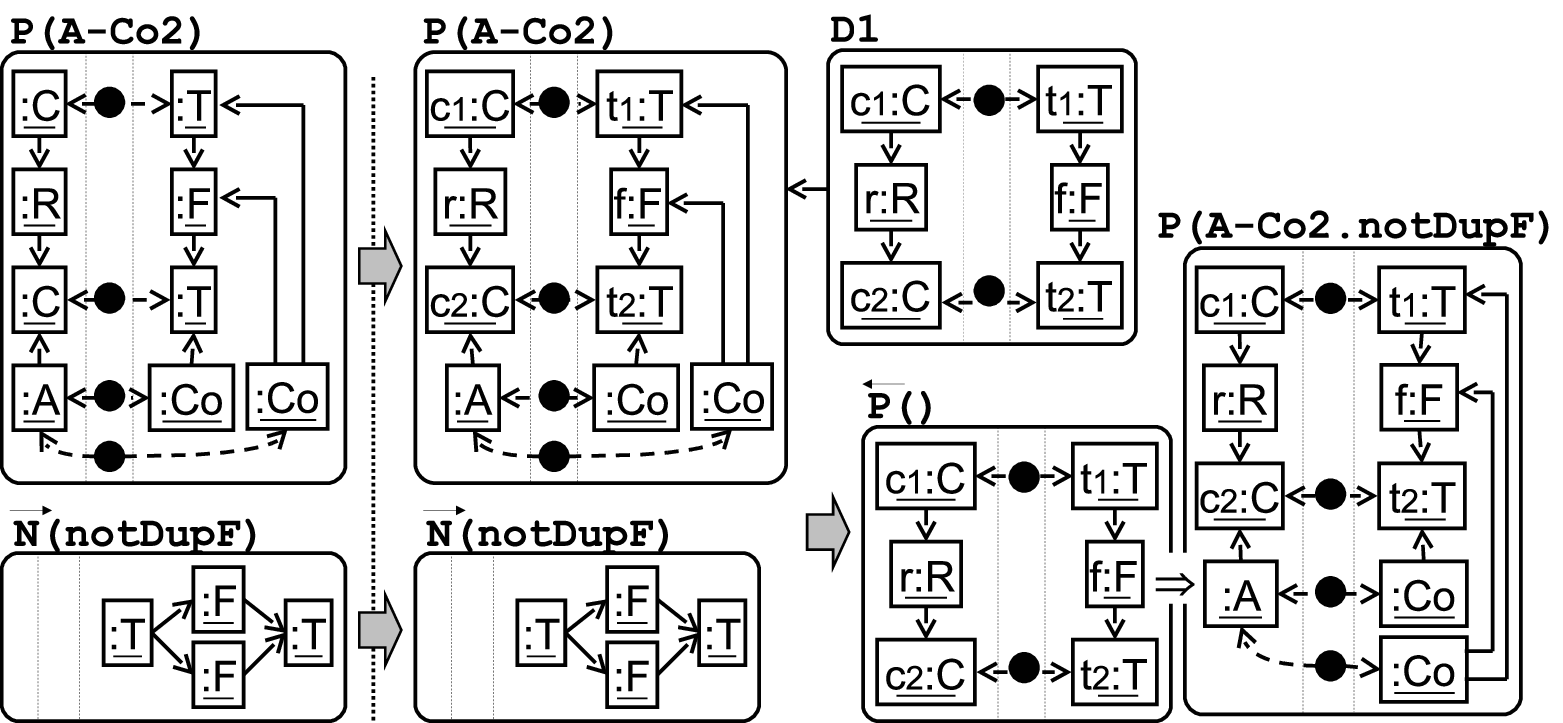}{0.4}{NP-Deduction and Annotation Example.}

\noindent {\bf Example.} Fig.~\ref{fig:NDed_example} shows the derivation of
C-Pattern \texttt{A-Co2.notDupF} from \texttt{A-Co2} and \texttt{notDupF}.
The latter is made of the pushout of two isomorphic graphs made of two $T$s and one $F$, which
belongs to \texttt{MIO(A-Co2, NotDupF)}. The completion of one of the isomorphic graphs
with respect to \texttt{A-Co2} is the pre-condition graph $\overleftarrow P()$ of \texttt{A-Co2.notDupF}. The newly
generated pattern reuses two $T$s and one $F$ so that the rule to be generated from it will not produce the situation
forbidden by \texttt{notDupF}. The annotation procedure adds a dependency
to \texttt{A-Co2} so that the generated rule will be mutually exclusive with the one
for the deduced pattern.

NP-Deduction 
has a generalization, called CNP-Deduction which handles the case of a C-Pattern and an N-Pattern.
The main difference with NP-Deduction is that the pre-condition of the generated pattern has to glue the positive 
pre-condition of the C-Pattern through the MIO (as in C-Deduction). 

\begin{prop}[CNP-Deduction]\label{prop:CNP_ded} $[\bigwedge_{i \in Pre}
\overleftarrow N(C_i) \wedge \overleftarrow{P}(C) \Rightarrow P(Q)] \wedge [\overrightarrow N(S)]$, with $S$ the pushout
of two isomorphic graphs $S_{1} \cong S_{2}$ and $S_1 \in MIO(Q, S)$ is equivalent 
to $[\bigwedge_{i \in Pre}
\overleftarrow N(C_i) \wedge \overleftarrow{P}(C) \Rightarrow P(Q)] \wedge [\overrightarrow N(S)] \wedge 
[\bigwedge_{i \in Pre}
\overleftarrow N(C_i) \wedge \overleftarrow P(P_s) \Rightarrow P(Q)]$, where $P_s$ is
calculated as shown in Fig.~\ref{fig:CNP}, with $cp \circ c = sp \circ b$ a P.O. square.
\end{prop}

\noindent {\em Proof.} In appendix. $\blacksquare$

\setlength{\unitlength}{0.7cm}
\begin{figure}[htb]
\centering
\makebox{\xymatrix@C=0.8cm@R=0.8cm{
    	     & S_{12}\ar[ld]\ar[rd]\ar@{}[dd]|{P.O.} &  &  & P_s\ar[rd]\\
S_2\ar[rd] &       & S_1\ar[ld]\ar[rd]\ar@{^{(}->}[d] &  &     & Q & C\ar[l]_{q}\ar@/_1pc/[llu]|{cp}\\
           & S     & C(S_1, Q)\ar@{^{(}->}[r]& Q &       &      C(S_1, Q)\ar@{^{(}->}[u]\ar@/^1pc/[uul]|{sp} & B_1\ar[l]|>>>>{b}\ar[u]|{c}\ar@{}[ul]|{P.B.}
 }}
\caption{CNP-Deduction.} \label{fig:CNP}
\end{figure}
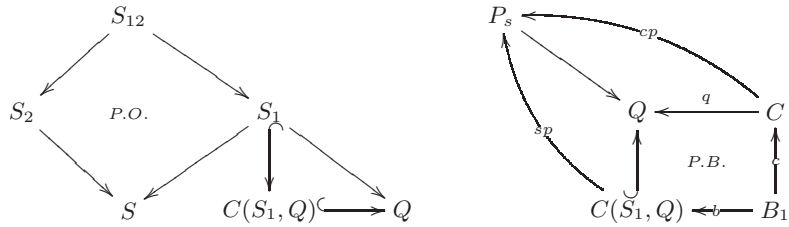

\section{Generating the Operational Rules}\label{sec:patterns}

This section details the generation of operational TGG rules from a M2M specification. Note that compilation
into other formalisms is possible, e.g. to a constraint satisfaction problem in the style of~\cite{ICMT}.
We first introduce the
structure of a non-deleting TGG rule.

\begin{defi}[Non-Deleting Oper. TGG Rule]\label{def:TGGGen}
A TGG rule $r=(L \overset{l} \rightarrow R, pre=\{\abb{n_i}{L}{N^i_L}\}_{i \in I}, post=\{\abb{n_j}{R}{N^j_R}\}_{j \in J})$ 
is made of an injective morphism $l$ of triple graphs, and sets $pre$ and $post$ of negative pre-
and post-conditions.
\end{defi}

Although negative post-conditions can be translated into
negative pre-conditions using the procedure in~\cite{Fundamentals,HeckelW95}, we use a set for them for simplicity
of presentation.

Next we show how to generate a TGG rule given an annotated C-Pattern. 
The main idea is to use $P_s= C +_{C|s} Q|_s = \langle Q|_s, C|_c, C|_t \rangle$ 
as the LHS (for the forward rule) and $Q$ as the RHS. The negative pre- and post-conditions of the C-Pattern are used as negative 
pre- and post-conditions of the rule. Note the similarities with the satisfiability of patterns (Def.~\ref{def:sat} and Fig.~\ref{fig:sat_diag}).
The rule's RHS is used as a negative pre-condition so that 
satisfiability is enforced only once.
Finally, dependencies are converted into negative pre-conditions. 

\begin{defi}[Derived TGG Rule] Given annotated pattern 
$T = ( \bigwedge_{i \in Pre} \overleftarrow N(C_i) \wedge \overleftarrow P(C) \Rightarrow P(Q) \bigwedge_{j \in Post} \overrightarrow N(C_j), 
D=\{\abb{n^k}{D_k}{Q}\}_{k \in K} ) $,
the following TGG operational rules are derived:
\begin{itemize}
\item {\bf Forward.} 
       $\overrightarrow{r_T}:\:(\srule{L=\langle Q|_s, C|_c, C|_t \rangle}{R=Q}{(id, q_c, q_t)}, 
         pre=\{\abb{n}{L}{R}\} \cup \{\abb{a^s_i}{L}{N^s_i} | L \ncong N^s_i\}_{i \in Pre} \cup \{\abb{s^k}{L}{S^k}\}_{k \in K}, 
         post=\{\abb{n_j}{R}{C_j}\}_{j \in Post})$. 
\item {\bf Backwards.} 
       $\overleftarrow{r_T}:\:(\srule{L=\langle C|_s, C|_c, Q|_t \rangle}{R=Q}{(q|_s, q|_c, id)}, 
         pre=\{\abb{n}{L}{R}\} \cup \{\abb{a^t_i}{L}{N^t_i} | L \ncong N^t_i\}_{i \in Pre} \cup \{\abb{s^k}{L}{S^k}\}_{k \in K}
	  , post=\{\abb{n_j}{R}{C_j}\}_{j \in Post})$.
\end{itemize}
\noindent where $N^x_i \cong C_i|_x +_{C|_x} C$, and $\abb{a^x_i}{L}{N^x_i}$ is uniquely determined (see 
	   Fig.~\ref{fig:sat_diag}, where $P_x=L$).
         $S^k$ is the left-extension of $D_k$, see left of 
         Fig.~\ref{fig:gen_rule}, where $n^k \circ b^k= r \circ l^k$ and 
         $d^k \circ b^k = s^k \circ l^k$ are pullback and pushout squares respectively.
\end{defi}

\setlength{\unitlength}{0.7cm}
\begin{figure}[htb]
\begin{picture}(12, 3.25)
\put(1,3)
{\xymatrix@C=0.5cm@R=0.5cm{
B_k \ar[r]^{b^k}\ar[d]^{l^k}\ar@{}[rd]|{P.B.} & D_{k} \ar[d]_{n^k}\ar@/^/[rdd]^{d^k} \\
L\ar@/_/[rrd]_{s^k}\ar[r]^>>>>{r}& R=Q\ar@{}[rd]|<<<{P.O.} \\
& & S^k
 }}

\put(7.5,0) {\includegraphics[scale=0.35]{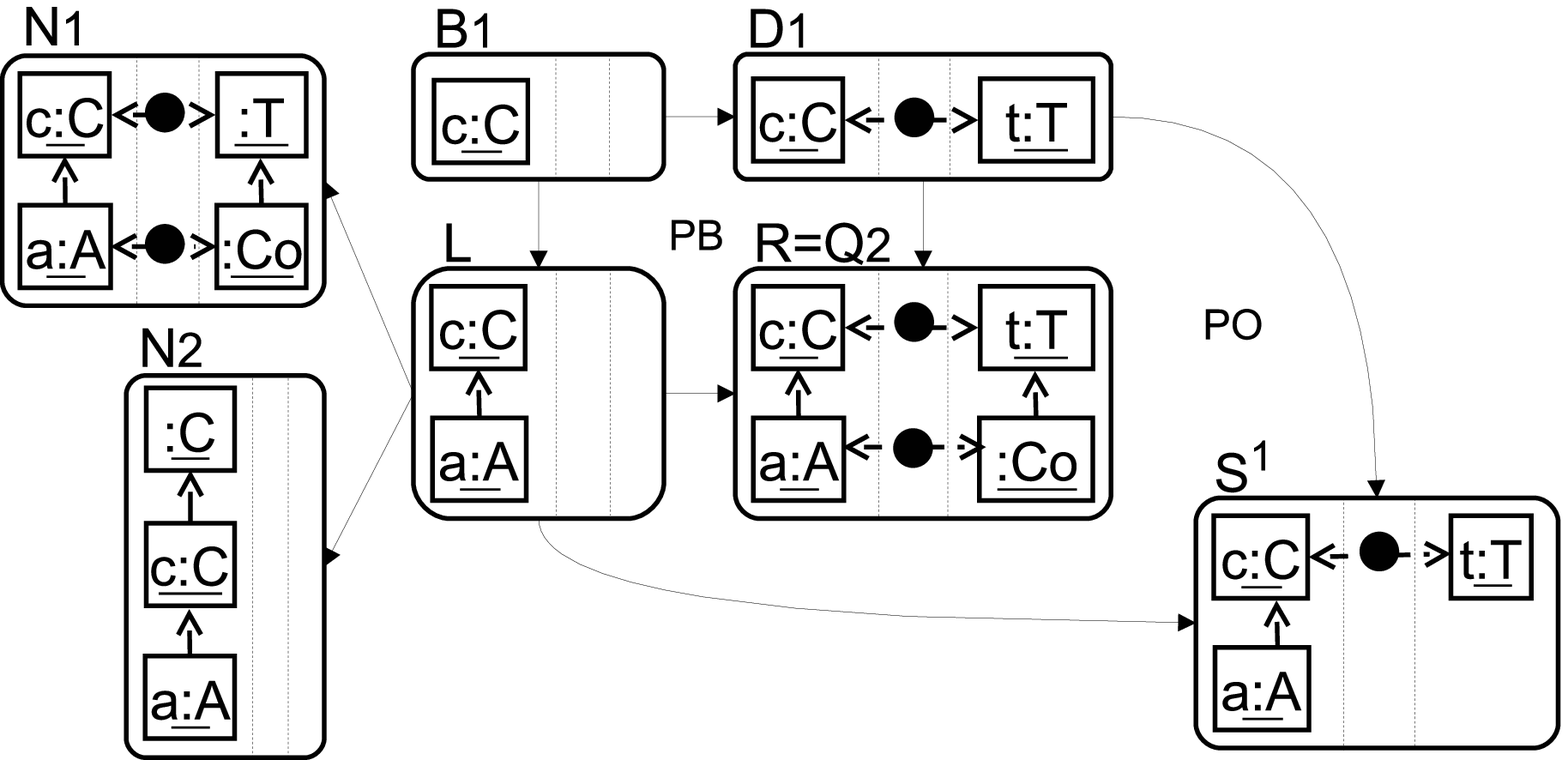}}
\end{picture}
\caption{Left Extension of $D_k \rightarrow Q$ (left). Generated Forward Rule \texttt{A-Co} (right).} \label{fig:gen_rule}
\end{figure}

\noindent {\bf Example.} The right of Fig.~\ref{fig:gen_rule} shows the generated forward rule from the annotated pattern \texttt{A-Co}
shown in Fig.~\ref{fig:S_Ded}. Note how the NAC $S^1$ forbids
applying the rule if the node $C$ has an associated $T$. In this case, the rule generated from the derived pattern
\texttt{C-T.A-Co} in Fig.~\ref{fig:S_Ded} would be applicable (see rule \texttt{C-T.A-Co} in Fig.~\ref{fig:rules}).

Before generating the rules we use the deduction and annotation
mechanisms on the initial M2M pattern specification in order to transform N-patterns into negative post-conditions of the 
other patterns, generate patterns that take into consideration the satisfaction of other patterns, and identify dependencies
between patterns. As stated before, we assume that the initial specification does not include patterns with both
a positive graph and a negative post-condition (as the latter can be expressed with N-Patterns).

\begin{defi}[Generation of Operational TGG Rules] Given specification $S$: 
\begin{enumerate}
\item Use PW (Def.~\ref{def:PW}) on all possible patterns.
\item Use C- or S-Annotation (Def.~\ref{def:SA}) for each pair of C- or S-Patterns. Do not derive a pattern if it
already exists.
\item Use NP-Annotation (Def.~\ref{def:NA}) on all possible patterns (initial and derived).
\item Use N-Deduction (Prop.~\ref{prop:N}) on all possible patterns and eliminate N-Patterns.
\item Take each derived pattern, and add to it all dependencies of the patterns it was derived from. 
Do not add such
dependencies if they are included in the positive pre-condition of the derived pattern, as the pattern would become useless.
\item Generate an operational TGG rule for each causal pattern (Def.~\ref{def:TGGGen}).
\end{enumerate}
\end{defi}

\figeps{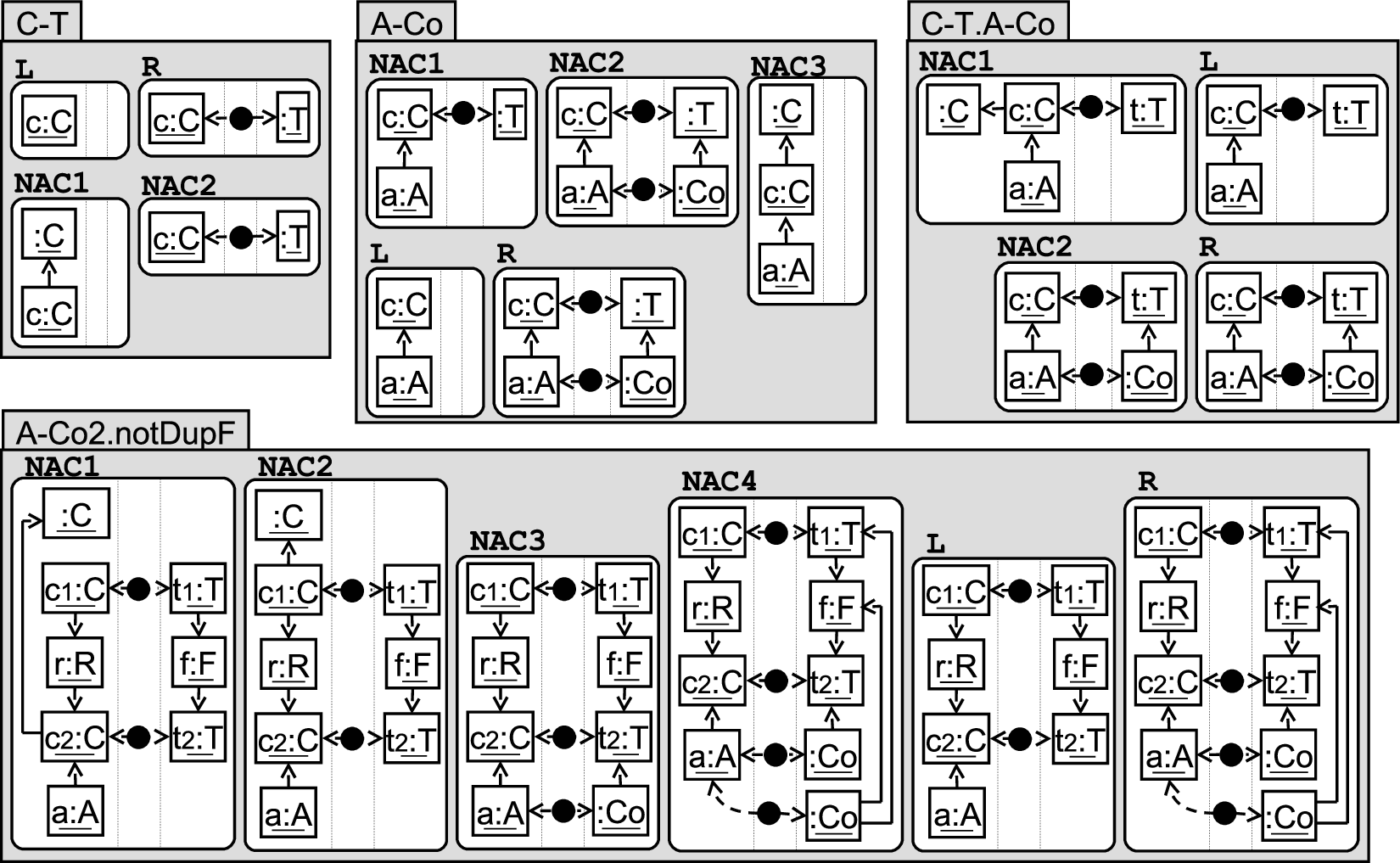}{0.37}{Some of the Generated Forward Operational Rules.}

\noindent {\bf Example.} Fig.~\ref{fig:rules} shows some of the generated forward rules. Rule \texttt{C-T}
is generated from pattern \texttt{C-T}. NAC1 results from a pre-condition, while NAC2
is equal to the RHS. Rule \texttt{A-Co} results from pattern \texttt{A-Co}. NAC3 comes
from the PW operation with pattern \texttt{C-T}, NAC2 is equal to RHS, and NAC1 is derived from a dependency
when making S-Deduction with \texttt{C-T}. Rule \texttt{C-T.A-Co} is generated from a pattern 
derived from \texttt{C-T} and \texttt{A-Co} through S-Deduction. Its first NAC comes from a dependency induced by their source patterns.
Finally, rule \texttt{A-Co.notDupF} results from NP-Deduction (see Fig.~\ref{fig:NDed_example}),
where NAC1 and NAC2 come from pre-conditions of the patterns from which it is derived, and NAC3 comes from a dependency. 
These two last rules have some additional NACs (not shown), stemming from N-Deduction with pattern \texttt{notDupF}.
The procedure generates a total of 10 rules, shown in Figs.~\ref{fig:rules2},~\ref{fig:rules3} and~\ref{fig:rules4}.

\figeps{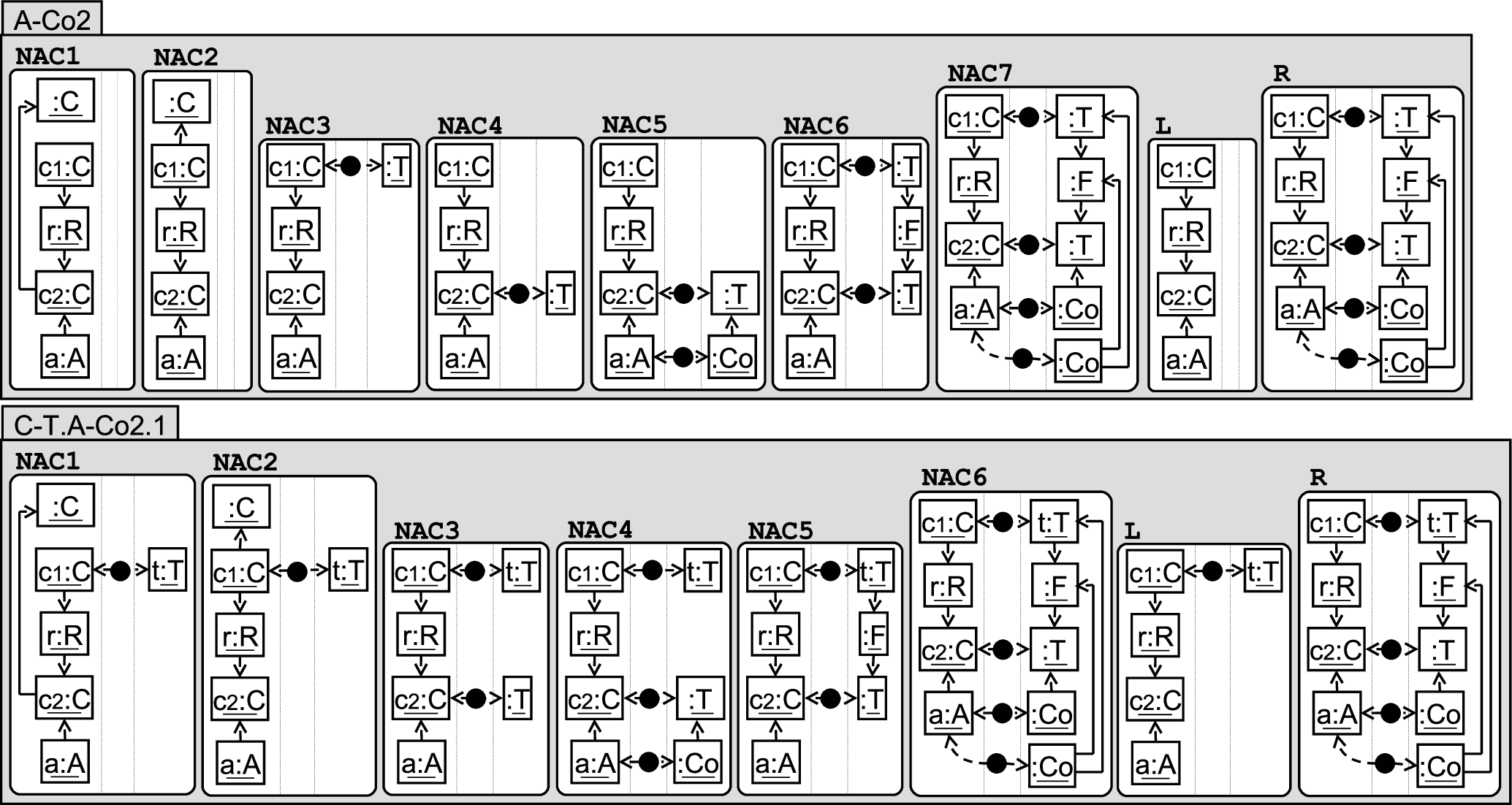}{0.35}{Some of the Generated Forward Operational Rules.}

\figeps{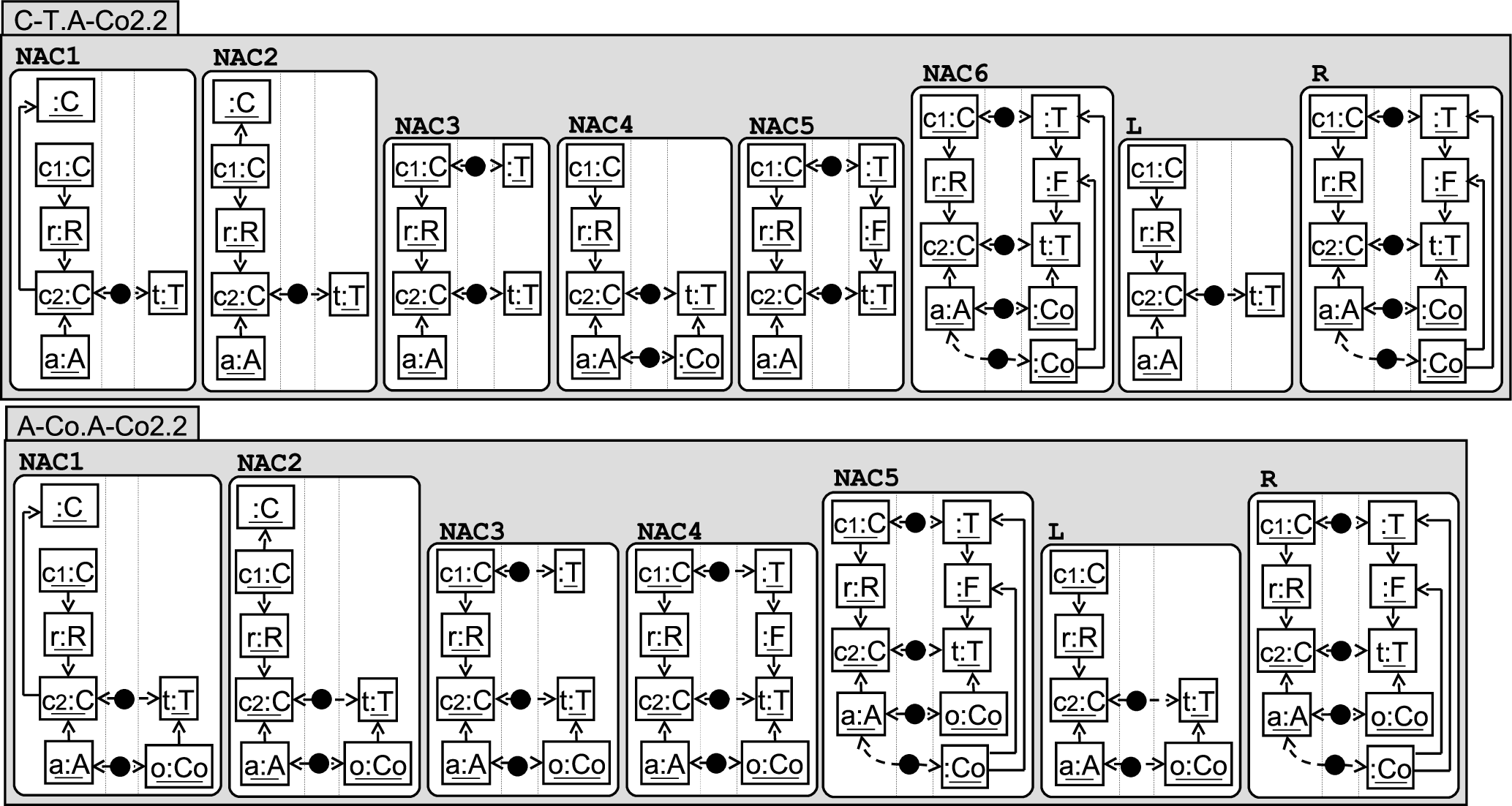}{0.35}{Some of the Generated Forward Operational Rules.}

\figeps{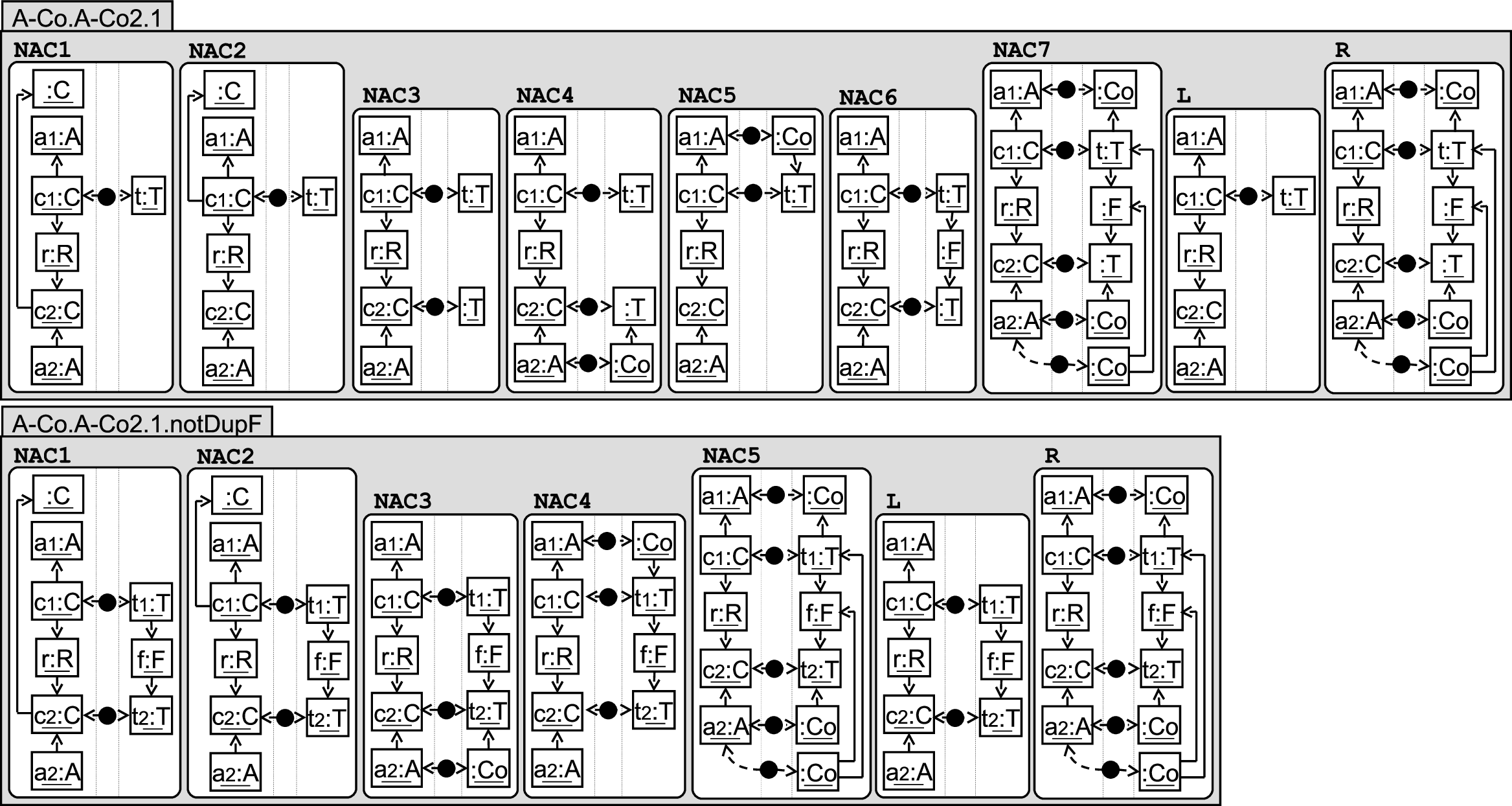}{0.33}{Some of the Generated Forward Operational Rules.}

\subsection{Correctness of the Operational Mechanisms}

Now we show the correctness of the generated rules, focussing on forward rules as a similar
reasoning holds for the backwards case. The generated rules: (i) must produce models satisfying the specification,
(ii) must be confluent, (iii) must terminate, and (iv) must transform each source model for which there is
a correct target model. 

\begin{itemize}
\item[(i)] follows from the construction of the TGG rules. Their LHS is 
$\langle Q|_s, C|_c, C|_t \rangle = C +_{C|_s} Q|_s =P_s$, which is the {\em base}
graph from which forward satisfaction is checked (see Fig.~\ref{fig:sat_diag}). As 
$R=Q$, morphism $\abb{m}{Q}{TrG}$ exists after the application of the rule. The rule negative 
pre- and post-conditions are derived from the negative pre- and post-conditions 
of the pattern. Thus, the rule can be applied iff the base morphism $m^s$ exists and the
negative pre- and post-conditions of the pattern are satisfied. The additional NAC $\cong$ R makes the rule enforce the pattern once.
As initially all forbidden graphs are expressed as N-Patterns, and we have performed N-Deduction, no
rule can produce a forbidden result. Since we start with an empty target graph, backwards satisfaction is
also obtained. Finally, the rule has additional NACs derived from dependencies, however
these just allow the execution of exactly one of the rules enforcing a given pattern so that they are confluent (see (ii)).

\item[(ii)] follows because S- and C-annotation add dependencies (which are transformed into NACs) to the initial patterns,
and these are appropriately propagated to their derived patterns in step 5 of the rule generation process. Fig.~\ref{fig:conf}
shows that a rule $L_1 \rightarrow R_1$ and a derived one through S-Deduction $L_3 \rightarrow R_3$ are mutually
exclusive. $S_1$ is the resulting NAC generated from the dependency of the first rule. In a situation where $L_1$
and $L_3$ are applicable, by the pushout universal property, there is a match of the NAC $S_1$. Thus, the first rule
is not applicable.

\setlength{\unitlength}{0.7cm}
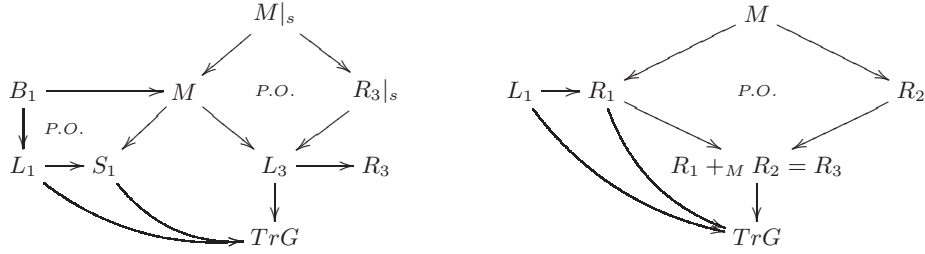
\begin{figure}[htb]
\centering
\makebox{\xymatrix@C=0.5cm@R=0.5cm{
    &     &   & M|_s\ar[ld]\ar[rd]\ar@{}[dd]|{P.O.} &  & & & & M\ar[ld]\ar[rd]\ar@{}[dd]|{P.O.}\\
B_1\ar[rr]\ar[d]\ar@{}[rd]|{P.O.} &     & M\ar[rd]\ar[ld] &      & R_3|_s\ar[ld] & & L_1\ar[r]\ar@/_1pc/[rrdd] & R_1\ar[rd]\ar@/_1pc/[rdd] & & R_2\ar[ld]\\
L_1\ar[r]\ar@/_1pc/[rrrd] & S_1\ar@/_1pc/[rrd] &   & L_3\ar[d]\ar[r]  & R_3  & & & & R_1+_M R_2=R_3\ar[d]\\
    &     &   & TrG & & & & & TrG
 }}
\caption{Mutual Exclusion of $L_1 \rightarrow R_1$ and $L_3 \rightarrow R_3$ (left). Applying first and third rule (right).} \label{fig:conf}
\end{figure}

Note however that initially we may have patterns included
in others: $[P(Q_1)] \wedge [P(Q_2)]$ with $Q_1 \hookrightarrow Q_2$. In this case,
S-Deduction generates $[P(Q_1)] \wedge [P(Q_2)] \wedge [P(Q_1) \Rightarrow P(Q_1 +_{Q_1} Q_2)]$ (assuming
just one MIO), from which we generate three rules. There is a conflict between the first two rules (i.e. a critical
pair). However in a situation where both the first and the second are applicable (e.g. if we have $Q_2|_s \hookrightarrow TrG$), applying 
the first and the third is equivalent to applying the second. 
The right of Fig.~\ref{fig:conf} shows that the existence of $R_1 \rightarrow TrG$ and $R_3 \rightarrow TrG$ implies the
existence of $R_2 \rightarrow TrG$.
Besides, we cannot apply the first and the second, because of the generated NACs: the second rule is added
$R_1$ as NAC (as $Q_1=R_1 \hookrightarrow Q_2=R_2$).

\noindent {\bf Example.} Consider the rules for the patterns \texttt{C-T} and \texttt{A-Co} and their
derived pattern (\texttt{C-T.A-Co} see Fig.~\ref{fig:rules}). Assume
a situation where both \texttt{C-T} and \texttt{A-Co} are applicable. If \texttt{C-T}
is applied first, then \texttt{A-Co} is disabled, but \texttt{C-T.A-Co} can be applied. If
\texttt{A-Co} is applied first, then no other rule is applicable. However, in both cases we reach the same
result.

\item[(iii)] follows from
the fact that (a) each rule has its RHS as a NAC, therefore it can only be applied once for each initial
match in the source model; and (b) a forward rule only changes the target model.

\item[(iv)] cannot be achieved for arbitrary M2M specifications. 
We restrict to what we call {\em Injective Positive Specifications}, which contain
enough positive patterns to produce the operational TGG rules. Next definition introduces the
forward case (FIP), the backwards one is similar.

\end{itemize}

\begin{figure}[htbp]
    \centering
    \subfigure{
        \includegraphics[scale = 0.47]{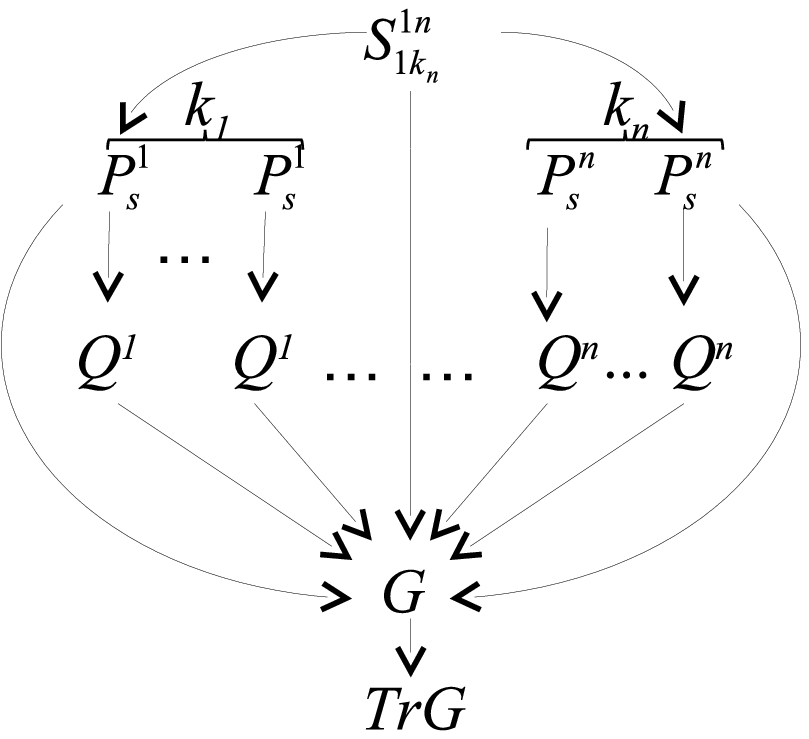}} \vline 
    \subfigure{
        \includegraphics[scale = 0.37]{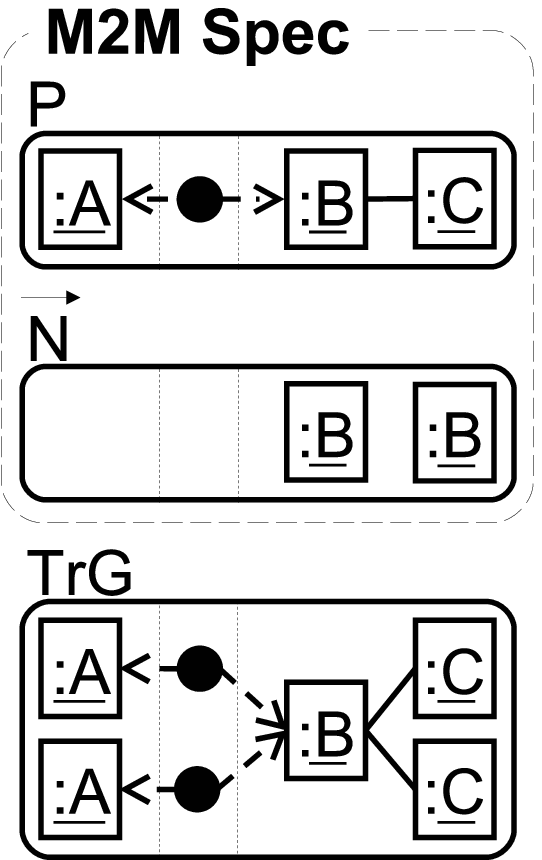}} \vline
    \subfigure{
        \includegraphics[scale = 0.37]{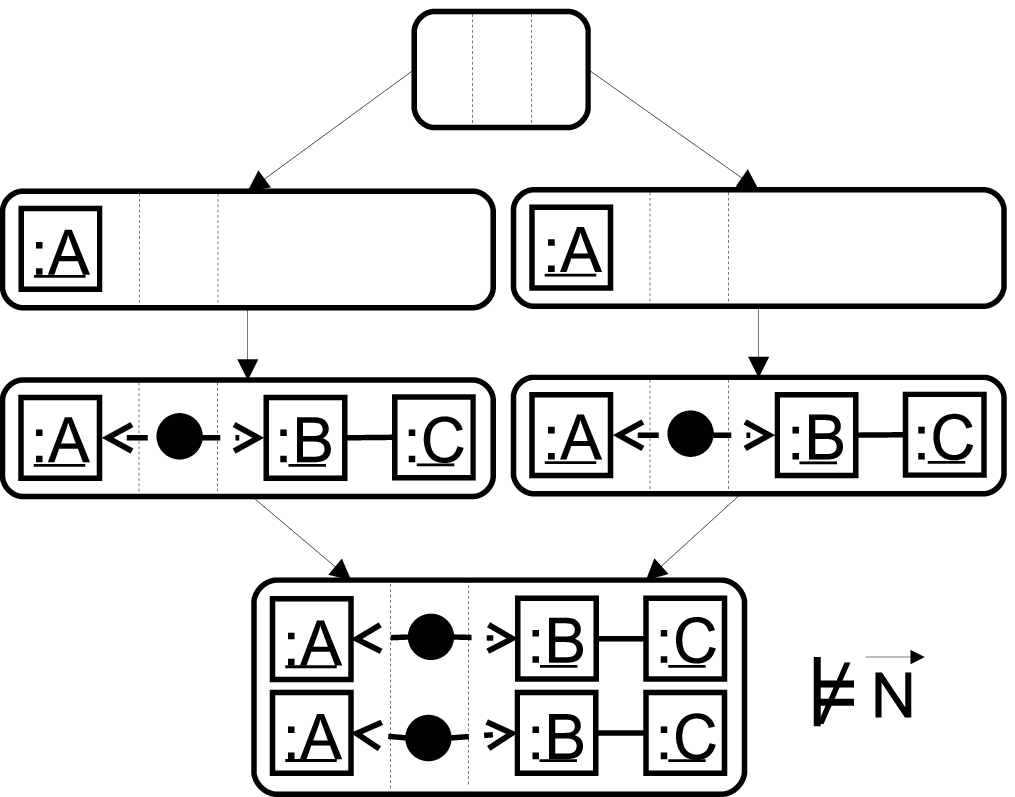}}
\caption{Condition for FIP (left). Non-FIP Specification (center). Invalid Graph (right).}\label{fig:FIP}
\end{figure}

\begin{defi}[FIP Spec.] Specification $S=\bigwedge_{i=1..n}T_i$ is FIP, 
iff $\forall M_s \in L(MM_s)$ s.t. $\exists TrG=\langle M_s, X \rangle \models S, \exists k_i  \in \mathbb N$,
$\exists P_s^i \leftarrow S^{ij}_{uv} \rightarrow P_s^j$ with $P^m_s = C^m+_{C^m|_s} Q^m|_s$, $u=\{1..k_i\}$,
$v=\{1..k_j\}$ and $S^{ij}_{uv} \ncong P^i_s$ if $i=j$, s.t. 
$G$ is the colimit of the diagram to the left of Fig.~\ref{fig:FIP} (with all arrows injective) with $G \hookrightarrow TrG$
and $G|_t \cong TrG|_t$.
\end{defi}

\noindent {\bf Remark.} The definition considers $k_i$ occurrences of each pattern $T_i$. Two occurrences of 
patterns $T_i$ and $T_j$ can overlap, and this is modelled by $S^{ij}_{uv}$. We forbid $P^i$ be
the overlap of two occurrences of the same pattern $Q^i$, as the operational mechanism minimally
enforces each pattern (i.e. rules have a NAC equal to the RHS). We have made a simplification in the diagram,
but each occurrence of $T_i$ should satisfy its negative pre- and post-conditions.

\begin{wrapfigure}{r}{0.35\textwidth}
\centering
\includegraphics[width=0.35\textwidth]{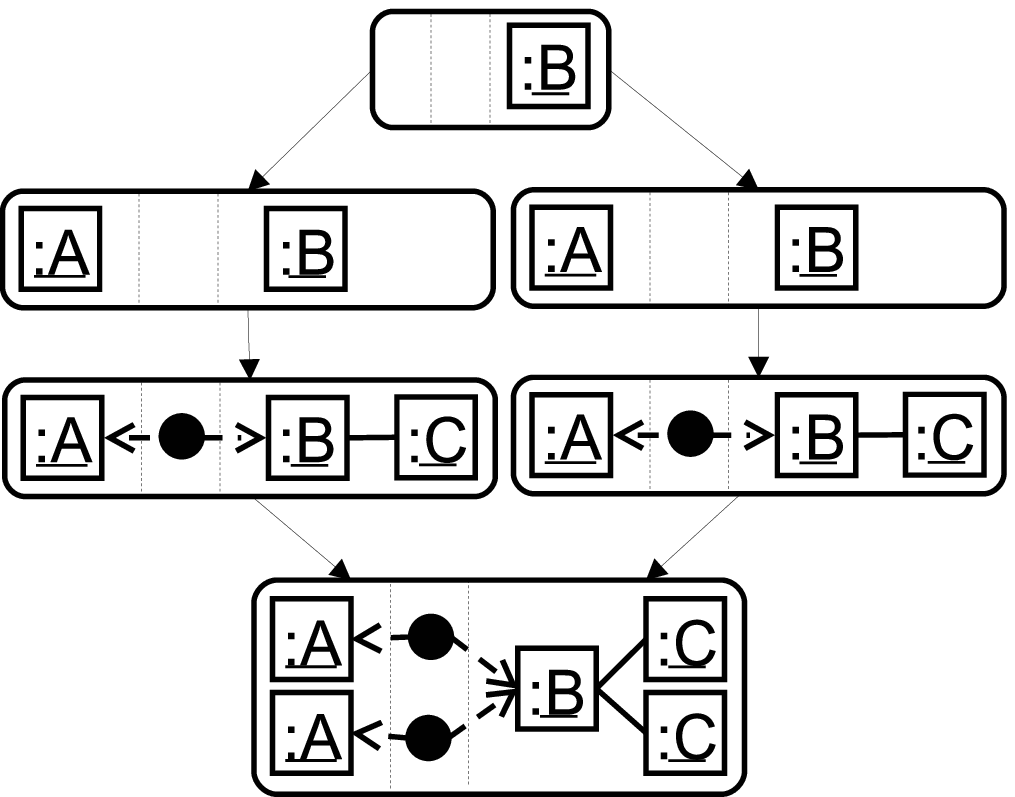}
\caption{FIP Specification.}
\label{fig:FIP2}
\noindent 
\end{wrapfigure}

\noindent {\bf Example.} Consider
the specification in the center of Fig.~\ref{fig:FIP}, and assume we do not perform any deduction. There is
a valid triple graph $TrG$ with two $As$ in its source, but the rules generated without deduction cannot
create such graph, as they would produce two $Bs$. 

The NP-Deduction rule can turn some non-FIP specifications into FIP. This is because it creates a new 
pattern that reuses an already created structure (in the target graph). The right of Fig.~\ref{fig:FIP}
shows that if NP-Deduction is not applied, we cannot handle a graph with two $As$. Fig.~\ref{fig:FIP2} shows
that after applying NP-Deduction, the resulting pattern can handle such graph as it reuses a $B$ and is
applied twice. It is up to future work to 
determine further deduction rules to cover additional non FIP-specifications.

\section{Analysis of the Transformation Specification}\label{sec:analysis}

This section gives an overview of some of the properties that
can be analysed from a pattern-based specification $S$. We distinguish three
levels: pattern, specifications and operational mechanisms.

\noindent {\bf Pattern Analysis.} We can analyse single patterns or pairs, for example:

\begin{itemize}
\item {\bf Pattern Conflict (PC).} Two patterns are in conflict if they express contradictory constraints.
This is for example the case of specification $[P(Q_P)] \wedge [\overrightarrow N(Q_N)]$ with $Q_N \hookrightarrow Q_P$.
In this case the generated TGG operational rule can never be applied. Thus, an initial 
graph of the form $TrG \cong Q_P|_s$ does not have a valid target model, and 
cannot be transformed according to $\overrightarrow S$. 

\item {\bf Tautology (T).} A pattern is a tautology if it is always satisfiable. This is the
case of patterns with the forms: (i) $[\overleftarrow N(Q) \Rightarrow P(Q)]$, as it is always either
vacuously or negatively satisfied, and (ii) $[\overleftarrow P(Q) \Rightarrow P(Q)]$ as it is always
either vacuously or positively satisfied.

\item {\bf Contradictory (C).} A pattern is contradictory if it can never be positively satisfied. This is
the case of a pattern with the form $[P(Q) \wedge \overrightarrow N(Q)]$, which can
only be vacuously satisfied.
\end{itemize}

\noindent {\bf Specification Analysis.} These define properties of the specification as a whole:

\begin{itemize}
\item {\bf Language Covering (LC).} A pattern specification is source covering iff 
$\biguplus_{P \in S^+} type(P|_{src})$ is surjective, where $S^+$ is the set of 
positive patterns of $S$. This means, that we have patterns that handle each 
construct of the source language. A similar notion can be defined for the target 
language. 

\item {\bf Full Forward, Backwards or Relating (FF, FB, FR).} A specification $S$ is FF (resp. FB)
iff $|Dom(\overrightarrow{S})| = |L(MM_s)|$ (resp. $|Dom(\overleftarrow{S})| = |L(MM_t)|$). A specification is
FR iff it is FF and FB.
If a specification is not FF, it means that is not defined for all valid models of the source language. 
Obviously, if a specification is not source LC it is neither FF nor FR (the converse in not necessarily true). 
Similarly a specification that is not target LC is not FB nor FR. A pattern conflict makes the specification
not to be FF or FB.

\item {\bf Contradiction(C).} A specification $S$ is a contradiction if the empty graph is the only graph able 
to satisfy it (i.e. no graph positively satisfies $S$). This is the case of a specification which is source
LC with all patterns contradictory.

\item {\bf Forward, Backwards or Relating Univalued (FU, BU, RU)}. $S$ is FU iff $\forall M_s \in Dom(\overrightarrow{S}),
|\{ M_T | \langle M_s, M_t \rangle \models S \}|=1$. Note that all specifications with our technique are FU, BU and RU.

\end{itemize}

\noindent {\bf Operational Mechanisms.} It is also possible to study properties of the generated TGG rules, for example:

\begin{itemize}

\item {\bf Hippocratic Transformation (HP).} (taken from~\cite{Stevens})
If $\forall M_{s} \in L(MM_s),$ $M_{t} \in L(MM_t)$ s.t. $\langle M_s, M_t \rangle \models S \Longrightarrow
[\overrightarrow{S}(\langle M_s, M_t \rangle)|_{t} \cong M_t \wedge  \overleftarrow{S}(\langle M_s, M_t \rangle)|_{s} \cong M_s]$. This 
property states that if a triple graph satisfies a specification, the application of $\overleftarrow S$ or $\overrightarrow S$
will not modify the triple graph. The transformations we generate are HP, as our rules minimally enforce each pattern
(they have NACs equal to RHS). A formal proof is left for future work.

\end{itemize}

\section{Related Work}\label{sec:related}

Some declarative approaches to M2M transformation use a textual 
syntax, e.g. PMT~\cite{PMT} or Tefkat~\cite{Tefkat}. 
These two particular notations are uni-directional, whereas we 
generate forward and backward transformations.

Among the visual declarative approaches, a prominent example is
the standard language QVT-relational~\cite{QVT}, which also has
a textual syntax. The relations 
may include {\em when} and {\em where} clauses that identify pre- 
and post-conditions and can refer to other relations. From this 
specification, executable QVT-core is generated that performs 
forward/backward transformations given a source/target model. 
This approach is similar to ours, but we
compile our patterns to TGG rules, allowing
the analysis of the generated transformation~\cite{Fundamentals}. Besides, 
we can make analysis at a higher-level as our patterns have a formal 
foundation. Moreover, we automatically 
detect pattern dependencies and perform pattern inference. In the QVT-relations language, dependencies must be made explicit 
in the {\em when} and {\em where} clauses, and there is no equivalent to our N-Patterns. An attempt to 
formalize QVT-core is found in~\cite{Greenyer06}.

In~\cite{Akehurst}, transformations are expressed through declarative
relations made of positive patterns, heavily relying on OCL constraints,
but no operational mechanism is given to enforce such relations.
In BOTL~\cite{BOTL}, the mapping rules use a UML-based 
notation that allows reasoning about applicability 
or meta-model conformance. We can reason both at the specification and 
operational levels.

TGGs~\cite{Schurr94} formalize the synchronized 
evolution of two graphs through declarative rules. From this specification, 
low-level operational TGG rules are derived to perform forward and 
backwards transformations, as well as to relate two existing graphs. 
We also generate these operational rules from 
our patterns. However, whereas in declarative TGG rules 
dependencies must be made explicit (i.e. we must say which elements should exist and which ones are created),
in our patterns this information is derived. For instance, in TGGs, a rule like
pattern \texttt{C-T.A-Co} has to be specified, it is not enough to give \texttt{C-T}
and \texttt{A-Co}. 

Although inspired by TGGs, our
patterns are a different approach to M2M transformation: patterns specify relations,
not rules. Similar to graph constraints~\cite{HeckelW95,Orejas}, a M2M specification by patterns describes 
a language of valid triple graphs.
Moreover, TGGs have some limitations. First, 
they do not allow specifying negative information,
nor deriving positive information from negative one (like NP-Deduction).
In~\cite{Konigs05}, the lack of negation
is alleviated by assigning execution priorities to rules. However, 
this is insufficient to simulate general application conditions, it has an operational 
nature, and implies knowing the rule generation mechanism and execution 
engine. Second, a control mechanism is needed to guide the execution of 
the operational rules, such as priorities~\cite{Konigs05} or their 
coupling to editing rules~\cite{EhrigTGG}. One can see TGGs as a subset
of our approach, where a TGG rule is a pattern of the form $\overleftarrow P(L) \Rightarrow P(R)$
without negative conditions or deduction techniques.

In~\cite{KindlerWagner}, an algorithm is given for the derivation of declarative TGGs from
example pairs of models. Interestingly, the user does not have to specify the correspondence
nodes in these pairs. The employed techniques resemble our use of MIOs, however, our patterns
are richer, as we allow negative pre- and post-conditions, and we have developed a theoretical
framework which includes further derivation techniques (e.g. NP-Deduction). 

With respect to graphical patterns, in the recent work in~\cite{Orejas}, a logic of constraints and some
deduction techniques were proposed. However, their basic constraints are existentially satisfied,
while ours are universal. Moreover, we provide deduction techniques
specially tailored for M2M specifications and triple patterns.
In~\cite{TriplePatterns} we presented a simpler notion of pattern 
and used it to extend normal rules to synchronous TGGs. We applied
it to the synchronization of the concrete and abstract syntax of visual models. 
The patterns were restricted to work with positive information, and the 
execution of the derived rules was associated to editing rules (like 
traditional TGGs). Here we present a new concept of pattern, which allows expressing negative conditions,
introduce deduction rules and present a new algorithm for TGG rule derivation that is suitable for 
M2M transformation and does not need a normal rule to start with.

\section{Conclusions and Future Work}\label{sec:conclusions}

In this paper we have presented a new formal approach to 
declarative M2M transformation. Relations between source 
and target models are expressed as different kinds of patterns, from
which operational TGG rules are derived
implementing forward/backwards
transformations and taking into account pattern 
interactions. This is done by deduction mechanisms that
detect interdependencies and 
produce new patterns that reuse structures created by other patterns.
This is one of the strengths of the present work: pattern dependencies
are automatically calculated and not explicitly given by the designer
such as with QVT and TGGs.
We have identified analysis properties, both at the specification
(e.g. language covering, pattern conflicts) and operational
levels (e.g. hippocratic transformations~\cite{Stevens}). 

Although we generate operational TGG rules 
from a pattern specification, other target formalisms 
could be used as well (e.g. OCL, Alloy). 
In fact, one of our next goals is expressing
a specification in terms of a constraint satisfaction problem,
in the lines of~\cite{ICMT}. This would eliminate some problems
of the compilation into rules, such as the restriction to handle FIP specifications
only. Note that with the theory presented so far we can handle
attributes, but not attribute conditions or computations. 
Our aim is to use OCL and the analysis techniques we proposed in~\cite{ICMT}.

We are also investigating additional analysis properties at the specification
and operational levels. 
It would be interesting to extend the set of 
derived operational rules to handle incremental 
synchronization and change propagation, using the techniques in~\cite{GuerraL06}. More complex
patterns able to deal with recursion or have parameters are also
under consideration. Finally, we aim to formalize
a part of QVT using this technique. 

{\bf Acknowledgements.} We thank the anonymous reviewers of ICGT'08 for their useful comments. 
Work supported by the Spanish Ministry of
Education and Science, projects TSI2005-08225-C07-06 and TIN2006-09678. 

\bibliographystyle{splncs}
\bibliography{article_ex}

\section*{Appendix: Proof of Some Results}

{\bf Proof of Prop.~\ref{prop:sp_ded}.} 
We show the equivalence of forward satisfiability, as a similar reasoning holds for the backwards one. 
We have to show that $TrG \models_F T_1=\bigwedge_{k \in \{1, 2\}} [\bigwedge_{i \in Pre^k} \overleftarrow{N}(C^k_i) \Rightarrow P(Q^k)]$
iff $TrG \models_F T_2= (\bigwedge_{k=\{1, 2\}} [\bigwedge_{i \in Pre^k} \overleftarrow{N}(C^k_i) \Rightarrow P(Q^k)])
\bigwedge_{M \in MIO(Q^1, Q^2)} [ \bigwedge_{i \in Pre^1 \cup Pre^2} \overleftarrow{N}(C'_i) \wedge
 \overleftarrow{P}(M) \Rightarrow P(Q^1 +_{M} Q^2)]$ .
If $TrG \models_F T_2$, obviously $TrG \models_F T_1$. 
Lets then assume that $TrG \models_F T_1$ and check all possible cases:

\begin{itemize}
\item[(a)] Both $P(Q^i)$ are positively satisfied and there is one occurrence of both: $\exists \abb{m^s_i}{Q^i|_s}{TrG}, \abb{m_i}{Q^i}{TrG}$ with
$m^s_i= m_i \circ q^s_i$ (see left of Fig.~\ref{fig:Positive_SAT}). There are two possibilities: either $\exists M_j \in MIO(Q^1, Q^2) \cong
m_1(Q^1) \times_{TrG} m_2(Q^2)$ or not. In the first case, 
$S^j=[\overleftarrow P(M_j) \Rightarrow P(Q^1 +_{M_j} Q^2)]$
is positively satisfied and the other $[\overleftarrow P(M_k) \Rightarrow P(Q^1 +_{M_k} Q^2)]$ are vacuously satisfied.
This is so as we first build the P.O. $Q^1+_{M_j} Q^2$, and then morphism $\abb{m^s_j}{Q^1+_{M_j} Q^2}{TrG}$ exists due to the
P.O. universal property (see left of Fig.~\ref{fig:Positive_SAT}). Moreover, $\abb{m_j}{P_s=M_j +_{M_j|_s} (Q^1+_{M_j} Q^2)|_s}{TrG}$
exists for the same reason (see right of Fig.~\ref{fig:Positive_SAT}).
In the second case, all patterns $[\overleftarrow P(M_k) \Rightarrow P(Q^1 +_{M_k} Q^2)]$ are vacuously satisfied.

\begin{figure}[htb]
\begin{picture}(12, 5)
\put(0,5)
{\xymatrix@C=0.5cm@R=0.5cm{
             &     & M_j\ar[ld]|{p^1}\ar[rd]|{p^2}\ar@{}[dd]|{P.O.}    & \\
Q^1|_s\ar[r]\ar@/_2pc/[rrdd]|{m^s_1} & Q^1\ar[rd]\ar@/_1pc/[rdd]|{m_1} & & Q^2\ar[ld]\ar@/^1pc/[ldd]|{m_2} & Q^2|_s \ar[l]\ar@/^2pc/[lldd]|{m^s_2} \\
             &     & Q^1+_{M_j} Q^2\ar[d]\ar[d]|{m^s_j} &\\
             &     & TrG\\
 }}

\put(11,5)
{\xymatrix@C=0.5cm@R=0.5cm{
             &  M_j|_s\ar[ld]\ar[rd]\ar@{}[dd]|{P.O.}    & \\
M_j\ar[rd]\ar@/_1pc/[rdd]|{m^1\circ p^1} &  & (Q^1+_{M_j} Q^2)|_s \ar@/^1pc/[ldd]|{m^s_j|_s} \ar[ld]\\
             &  P_s\ar[d]|{m_j} \\
             &  TrG\\
 }}
\end{picture}
\caption{Positive Satisfaction of $P(Q^1+_{M_j} Q^2)$ given positive satisfaction of $P(Q^i)$.}\label{fig:Positive_SAT}
\end{figure}
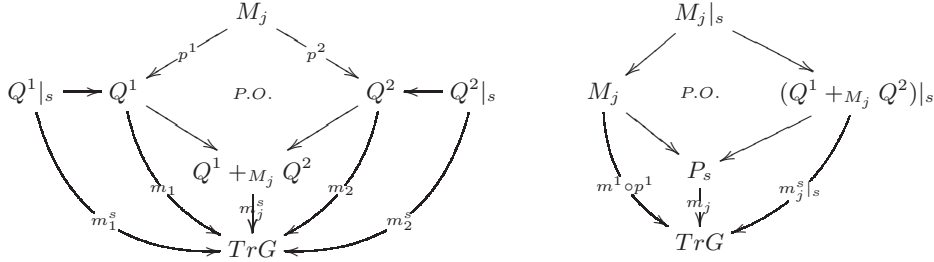

\item[(b)] $P(Q^1)$ is positively satisfied and $P(Q^2)$ vacuously satisfied. In this case $\nexists \abb{m^s_2}{Q_2|_s}{TrG}$,
and therefore $\nexists \abb{m_k}{M_k +_{M_k|_s} ((Q^1 +_{M_k} Q^2)|_s)}{TrG}$ (because $Q^2|_s \hookrightarrow 
M_k +_{M_k|_s} ((Q^1 +_{M_k} Q^2)|_s)$) and therefore each $\overleftarrow P(M_k) \Rightarrow P(Q^1 +_{M_k} Q^2)$ is vacuously satisfied.

\item[(c)] If both $P(Q^i)$ are vacuously satisfied by the same reasoning each $\overleftarrow P(M_k) \Rightarrow P(Q^1 +_{M_k} Q^2)$ is vacuously 
satisfied.

\item[(d)] If $P(Q^1)$ is positively satisfied and $P(Q^2)$ is negatively satisfied (assume just one instance of each), 
then $\exists \abb{m^s_1}{Q^1|_s}{TrG}, \abb{m^s_2}{Q^2|_s}{TrG}, \abb{m_1}{Q^1}{TrG},
(C^2_i|_s=N^{2s}_i)\overset{n^s_i} \rightarrow {TrG}$ (see Fig.~\ref{fig:sat_diag}). There are two possibilities: either 
$\exists M_j \in MIO(Q^1, Q^2)$ s.t. $M_j|_s \cong m^s_1(Q^1|_s) \times_{TrG} m^s_2(Q^2|_s)$ or not. In the
second case, all $\overleftarrow P(M_k)$ are vacuously satisfied. 

\begin{figure}[htb]
\begin{picture}(12, 5)
\put(0,5)
{\xymatrix@C=0.5cm@R=0.7cm{
             &     & M_j|_s\ar[ld]|{p^1_s}\ar[rd]|{p^2_s}\ar@{}[dd]|{P.B.}    & \\
Q^1\ar@/_1pc/[rrd]|{m_1} & Q^1|_s\ar[l]\ar@/_1pc/[rd]|{m^s_1} & & Q^2|_s\ar@/^1pc/[ld]|{m^s_2}\ar[r] & C^2_i|_s\ar@/^1pc/[lld]|{n^s_i} \\
             &     & TrG\\
 }}

\put(11,5)
{\xymatrix@C=0.5cm@R=0.5cm{
             &  M_j|_s\ar[ld]\ar[rd]\ar@{}[dd]|{P.O.}    & \\
Q^1|_s\ar[rd]\ar@/_1pc/[rdd]|{m^s_1} &  & Q^2|_s \ar@/^1pc/[ldd]|{m^s_2} \ar[ld]\\
             &  Q^1|_s+_{M_j|_s} Q^2|_s \ar[d]|{e_j} \\
             &  TrG\\
 }}
\end{picture}
\caption{Positive Satisfaction of $P(M_j)$ given positive satisfaction of $P(Q^1)$ and negative satisfaction of $P(Q^2)$.}\label{fig:Negative_SAT}
\end{figure}

In the first case we have that:
\begin{itemize}
\item $\exists \abb{e_j}{Q^1|_s +_{M_j|_s} Q^2|_s = (Q^1+_{M_j} Q^2)|_s}{TrG}$ due to the P.O. universal property. See the right of
Fig.~\ref{fig:Negative_SAT}.
\item $\exists \abb{m^s_j}{M^s_j= M_j +_{M_j|_s} Q^1 +_M Q^2|_s}{TrG}$ due to the P.O. universal property, thus the base morphism of $M_j$ exists.
See the left of Fig.~\ref{fig:Negative_SAT2}. In this diagram, square $M_j|_s \hookrightarrow M_j \rightarrow Q_1 \rightarrow M'^s_j = M_j|_s \rightarrow
(Q^1 +_{M_j}Q^2)|_s \rightarrow M'^s_j$ is a P.O. Then, $m^s_j= m''^s_j \circ m'^s_j$.
\item Morphism $\abb{f_i}{C'_i|_s}{TrG}$ (where $C'_i|_s$ is the source component of the 
pre-condition of $M_j$ derived from $C^2_i$) exists due to the
universal P.O. property, see the right of Fig.~\ref{fig:Negative_SAT2}. 
Therefore $M_j$ is negatively satisfied. The other $P(M_k)$ are vacuously satisfied.
\end{itemize}

\begin{figure}[htb]
\begin{picture}(12, 7)
\put(0,7)
{\xymatrix@C=0.4cm@R=0.7cm{
    &     & M_j|_s\ar@{^{(}->}[ld]\ar[rd]\ar@{}[dd]|{P.O.} & \\
    & M_j\ar[ld]\ar[rd] &                      & Q^1+_{M_j}Q^2|_s\ar[ld]\ar[ldd]\ar@/^1pc/[lddd]\\
Q^1\ar[rrd]\ar@/_1pc/[rrdd] &     & M^s_j\ar[d]|{m'^s_j}& \\
    &     & M'^s_j\ar[d]|{m''^s_j} &\\
    &     & TrG 
 }}

\put(8,7)
{\xymatrix@C=0.4cm@R=0.5cm{
             &  M_j|_s\ar[ld]\ar[rd]\ar@{}[dd]|{P.O.}      & \\
Q^1|_s\ar[rd]&                                             & Q^2|_s \ar[ld]\ar[r]\ar@{}[d]|{P.O.} & C^2_i|_s\ar[ld]\ar@/^2pc/[lldd]|{n^s_i}\\
             &  Q^1|_s+_{M_j|_s} Q^2|_s \ar[d]|{e_j}\ar[r] &  C'_i|_s\ar@/^1pc/[ld]|{f_i} \\
             &  TrG\\
 }}
\end{picture}
\caption{Positive Satisfaction of $P(M_j)$ given positive satisfaction of $P(Q^1)$ and negative satisfaction of $P(Q^2)$.}\label{fig:Negative_SAT2}
\end{figure}

\item[(e)] A similar reasoning follows if both $P(Q^i)$ are negatively satisfied.
\item[(f)] If $P(Q^1)$ is negatively satisfied (with just one instance) and $P(Q^2)$ is vacuously satisfied, then all
$P(M_j)$ are vacuously satisfied.

\end{itemize}

If there is more than one occurrence of a pattern either positively or negatively satisfied, we do all combinations
but they reduce to the previous cases. Now let's assume that some $T_i$ is not satisfied. If $TrG \nvDash_F T_1$, obviously $TrG \nvDash_F T_2$. Let's
assume that $TrG \nvDash_F T_2$. We check all the cases:

\begin{itemize}
\item[(a)] If $TrG \nvDash_F P(Q^i)$, then $TrG \nvDash T_1$.
\item[(b)] If $TrG \nvDash_F \overleftarrow P(M_j) \Rightarrow P(Q^1 +_{M_j} Q^2)$, this means that $\exists \abb{m^s_j}{M_j +_{M_j|_s} ((Q^1 +_{M_j} Q^2)|_s)}{TrG}$,
$\nexists \abb{m_j}{Q^1 +_{M_j} Q^2}{TrG}$ and $\nexists \abb{n'^s_u}{N'_u}{TrG}$ (where $N'_u$ are the derived negative pre-conditions).
As $m^s_j$ exists, both $\abb{m^s_i}{Q^i|_s}{TrG}$ exist
because $Q^i|_s \hookrightarrow M_j +_{M_j|_s} ((Q^1 +_{M_j} Q^2)|_s)$. As no $n'^s_u$ exists, the negative pre-conditions of $Q^i$
are satisfied. As
$m_j$ does not exist, some $\abb{m_i}{Q^i}{TrG}$ must not exist because $Q^i \hookrightarrow Q^1 +_{M_j} Q^2$.
Thus, either $TrG \nvDash_F P(Q^1)$ or $TrG \nvDash_F P(Q^2)$, therefore $TrG \nvDash_F T_1$.
\end{itemize}
$\blacksquare$


{\bf Proof of Prop.~\ref{prop:cp_ded}.} The proof is similar to the one for S-Deduction.
We show the equivalence of forward satisfiability, as a similar reasoning holds for the backwards one. 
We have to show that $TrG \models_F T_1=\bigwedge_{k \in \{1, 2\}} [\bigwedge_{i \in Pre^k} \overleftarrow{N}(C^k_i) \wedge \overleftarrow{P}(C^k) \Rightarrow P(Q^k)]$
iff $TrG \models_F T_2= (\bigwedge_{k \in \{1, 2\}} [\bigwedge_{i \in Pre^k} \overleftarrow{N}(C^k_i) \wedge \overleftarrow{P}(C^k) \Rightarrow P(Q^k)])
\bigwedge_{M \in MIO(Q^1, Q^2)} [ \bigwedge_{i \in Pre^1 \cup Pre^2} \overleftarrow{N}(C'_i) \wedge
 \overleftarrow{P}(P^c) \Rightarrow P(Q^1 +_{M} Q^2)]$ (see Fig.~\ref{fig:C_Ded}).
If $TrG \models_F T_2$, obviously $TrG \models_F T_1$. 
Lets then assume that $TrG \models_F T_1$ and check all possible cases:

\begin{itemize}
\item [(a)] Both $P(Q^i)$ are positively satisfied and there is one occurrence of both: $\exists \abb{m^s_i}{P^i_s}{TrG}, \abb{m_i}{Q^i}{TrG}$ with
$m^s_i= m_i \circ q^s_i$. There are two possibilities: either $\exists M_j \in MIO(Q^1, Q^2) \cong
m_1(Q^1) \times_{TrG} m_2(Q^2)$ or not. In the second case, all $S^k=[\overleftarrow P(P^c_k) \Rightarrow P(Q^1 +_{M_k} Q^2)]$
are vacuously satisfied.
In the first case, 
$S^j=[\overleftarrow P(P^c_j) \Rightarrow P(Q^1 +_{M_j} Q^2)]$
is positively satisfied and the other $[\overleftarrow P(P^c_k) \Rightarrow P(Q^1 +_{M_k} Q^2)]$ are vacuously satisfied.
This is so as (i) $\exists Q^1 +_{M_j} Q^2$ due to the P.O. universal property, (ii) $\exists P^c \rightarrow TrG$ for the same
reason and (iii) $\exists M^s_j \rightarrow TrG$.

\item[(b)] $P(Q^1)$ is positively satisfied and $P(Q^2)$ vacuously satisfied. In this case $\nexists \abb{m^s_2}{P^2_s}{TrG}$,
and therefore $\nexists \abb{m^c_j}{P_j^c +_{P_j^c|_s} ((Q^1 +_{M_j} Q^2)|_s)}{TrG}$ (because $Q^2|_s \hookrightarrow 
P^c_j +_{P^c_j|_s} ((Q^1 +_{M_j} Q^2)|_s)$) and therefore each $\overleftarrow P(P^c_k) \Rightarrow P(Q^1 +_{M_k} Q^2)$ is vacuously satisfied.

\item[(c)] If both $P(Q^i)$ are vacuously satisfied by the same reasoning each $\overleftarrow P(P^c_k) \Rightarrow P(Q^1 +_{M_k} Q^2)$ is vacuously 
satisfied.

\item[(d)] If $P(Q^1)$ is positively satisfied and $P(Q^2)$ is negatively satisfied (assume just one instance of each), 
then either all $\overleftarrow P(P^c_k)$ are vacuously satisfied, or one $\overleftarrow P(P^c_j)$ is negatively satisfied.

\item[(e)] A similar reasoning follows if both $P(Q^i)$ are negatively satisfied.

\item[(f)] If $P(Q^1)$ is negatively satisfied (with just one instance) and $P(Q^2)$ is vacuously satisfied, then all
$P(P^c_j)$ are vacuously satisfied.

If there is more than one occurrence of a pattern either positively or negatively satisfied, we do all combinations
but they reduce to the previous cases. If some $T_i$ is not satisfied the proof is analogous to the one for S-Deduction.
 
\end{itemize}

$\blacksquare$


{\bf Proof of Prop.~\ref{prop:CNP_ded}.} We have to proof that if $T_1 = \overleftarrow P(C) \Rightarrow P(Q)$ is
satisfied, so is $T_2 = \overleftarrow P(T_s) \Rightarrow P(Q)$ with $C \hookrightarrow T_s \hookrightarrow Q$. Again,
we show forward satisfaction, as a similar reasoning holds for the backwards one.

If $TrG \models_F T_2$, then $TrG \models_F T_1$. Let's assume that $TrG \models_F T_1$ is satisfied and check
all the cases:

\begin{itemize}
\item[(a)] If $T_1$ is positively satisfied, then so is $T_2$, as Fig.~\ref{fig:CNP_pos_sat} shows. The left
diagram shows that $\exists T_s \rightarrow TrG$ due to the P.O. universal property. The right diagram
shows that $\exists P^2_s = T_s +_{T_s|_s} Q|_s \rightarrow TrG$ for the same reason.

\begin{figure}[htb]
\begin{picture}(12, 5)
\put(0,5)
{\xymatrix@C=0.4cm@R=0.7cm{
    & C|_s\ar@{^{(}->}[ld]\ar[rd]\ar@{}[dd]|{P.O.} & & & B_1\ar[ld]\ar[rd]\ar@{}[dd]|{P.O.}\\
C\ar[rd] &                      & Q|_s\ar[ld] & C(S, Q)\ar[rd]\ar[d]& & C\ar[ld]\ar@/^2pc/[lldd]\\
& P^1_s\ar[rr]\ar@/_1pc/[rrd] & & Q\ar[d] & T_s\ar[l]\ar@/^1pc/[ld]\\
 &     & & TrG 
 }}

\put(11,5)
{\xymatrix@C=0.4cm@R=0.5cm{
    & T_s|_s\ar[ld]\ar[rd]\ar@{}[dd]|{P.O.}\\
T_s\ar[rd]\ar@/_1pc/[rdd] & & Q|_s\ar[ld]\ar@/^1pc/[ldd]\\
    & P^2_s\ar[d] \\
    & TrG
 }}
\end{picture}
\caption{Positive satisfaction of $T_2$ given positive satisfaction of $T_1$.}\label{fig:CNP_pos_sat}
\end{figure}
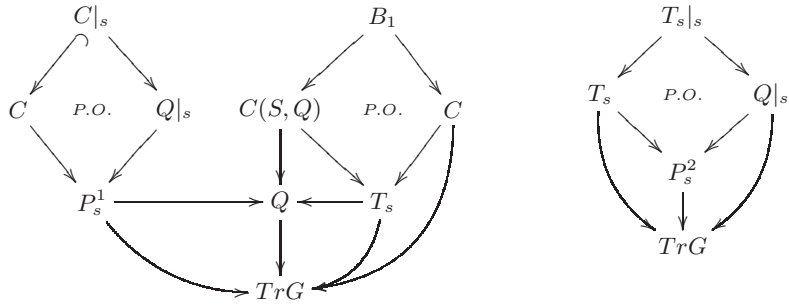

\item[(b)] Lets assume that $T_1$ is negatively satisfied. Then $\exists C_i |_s \rightarrow N_i \rightarrow TrG$.
There are two options, either $\exists Q \rightarrow TrG$ or not. In the first case, by item (a), we have that $\exists T_s \rightarrow TrG$.
But then, by the P.O. universal property $\exists N'^i \rightarrow TrG$, and hence $T_2$ is negatively satisfied.
If $\nexists Q \rightarrow TrG$, then either $\exists P^2_s \rightarrow TrG$ or not. In the first case $\exists T_s \rightarrow TrG$
and therefore $\exists N'^i \rightarrow TrG$ and $T_2$ is negatively satisfied. In the second case $T_2$ is vacuously satisfied.
See the left of Fig.~\ref{fig:CNP_neg_sat}.

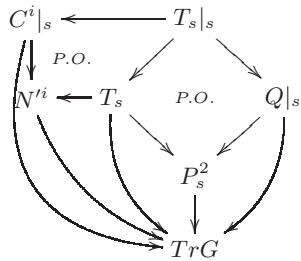
\begin{figure}[htb]
\begin{picture}(12, 5)
\put(5,4)
{\xymatrix@C=0.4cm@R=0.5cm{
C^i|_s\ar[d]\ar@{}[rd]|{P.O.}\ar@/_3pc/[rrddd]&    & T_s|_s\ar[ll]\ar[ld]\ar[rd]\ar@{}[dd]|{P.O.}\\
N'^i\ar@/_1pc/[rrdd] & T_s\ar[l]\ar[rd]\ar@/_1pc/[rdd] & & Q|_s\ar[ld]\ar@/^1pc/[ldd]\\
&    & P^2_s\ar[d] \\
 &   & TrG
 }}
\end{picture}
\caption{Satisfaction of $T_2$ given negative satisfaction of $T_1$ (left).}\label{fig:CNP_neg_sat}
\end{figure}

\item[(c)] Lets assume that $T_1$ is vacuously satisfied. Then (i) $\nexists T_s \rightarrow TrG$, because, as
$\exists C \rightarrow T_s$ if it would exist then we would have $C \rightarrow TrG$; (ii) $\nexists P^2_s \rightarrow TrG$,
as $\nexists T_s \rightarrow TrG$ and $T_s \rightarrow P^2_s$. Hence, $T_2$ is vacuously satisfied.

\end{itemize}
$\blacksquare$

\end{document}